\documentclass[lettersize,journal]{IEEEtran}
\usepackage{amsmath,amsfonts, amssymb, amsthm}
\usepackage{algorithmic}
\usepackage{setspace}
\usepackage[ruled,lined]{algorithm2e}

\usepackage{multicol}
\usepackage{amsthm}

\newtheorem{mydef3}{Lemma}

\usepackage{mathrsfs}  
\usepackage{graphicx}
\usepackage{xcolor}
\usepackage{blindtext}

\usepackage{graphicx,epsf,epic,eepic,psfig,latexcad}
\renewcommand{\baselinestretch}{1}

\usepackage{amssymb,amsmath,caption}
\usepackage{cite}
\usepackage{amsmath, amssymb, amsthm}
\usepackage{graphicx}
\renewcommand{\baselinestretch}{1}

\usepackage{epstopdf}

\DeclareMathOperator*{\argmin}{arg\,min} % thin space, limits underneath in displays
\DeclareMathOperator*{\argmax}{arg\,max} % thin space, limits underneath in displays
\newcommand\numeq[1]%
  {\stackrel{\scriptscriptstyle(\mkern-1.5mu#1\mkern-1.5mu)}{=}}
\makeatletter
\usepackage{textcomp}
\makeatother
\usepackage{xcolor}
\begin{document}
	\title{RIS-Aided Unsourced Multiple Access (RISUMA): Coding Strategy and Performance Limits}
	\author{Mohammad Javad Ahmadi, Mohammad Kazemi, and Tolga M. Duman \vspace{-2mm}	
\thanks{M. J. Ahmadi is with the Chair of Information Theory and Machine Learning, Technische Universität Dresden, 01062
Dresden, German (E-mail: mohammad\_javad.ahmadi@tu-dresden.de), M. Kazemi is with the Department of Electrical and Electronic Engineering, Imperial College London, London SW7 2BT, U.K. (e-mail: mohammad.kazemi@imperial.ac.uk), T. M. Duman is with the Department of Electrical and Electronics Engineering, Bilkent University, Ankara, 06800, Turkey (email: duman@ee.bilkent.edu.tr)}
 \thanks{This research is funded by the Scientific and Technological Research Council of Turkey (TUBITAK) under the grant 119E589. Mohammad Kazemi’s work was partly funded by UK Research and Innovation (UKRI) under the UK government’s Horizon Europe funding guarantee [grant number 101103430]. This paper was presented in part at the 2023 IEEE Global Communications Conference (GLOBECOM).}
 %\thanks{The authors are with the Department of Electrical and Electronics Engineering, Bilkent University, 06800 Ankara, Turkey (e-mails: \{ahmadi, kazemi, duman\}@ee.bilkent.edu.tr).}
	}
\maketitle
\begin{abstract}
This paper considers an unsourced random access (URA) set-up equipped with a passive reconfigurable intelligent surface (RIS), where a massive number of unidentified users (only a small fraction of them being active at any given time) are connected to the base station (BS). We introduce a slotted coding scheme for which each active user chooses a slot at random for transmitting its signal, consisting of a pilot part and a randomly spread polar codeword. The proposed decoder operates in two phases. In the first phase, called the RIS configuration phase, the BS detects the transmitted pilots. The detected pilots are then utilized to estimate the corresponding users' channel state information, using which the BS suitably selects RIS phase shift employing the proposed RIS design algorithms. The proposed channel estimator offers the capability to obtain the channel coefficients of the users whose pilots interfere with each other without prior access to the list of transmitted pilots or the number of active users. In the second phase, called the data phase, transmitted messages of active users are decoded. Moreover, we establish an approximate achievability bound for the RIS-based URA scheme, providing a valuable benchmark. Computer simulations show that the proposed scheme outperforms the state-of-the-art for RIS-aided URA.
\end{abstract}

\section{Introduction}
Unsourced random access (URA) is an important paradigm to support applications such as Internet-of-things (IoT), where a large number of users with small payloads share a common codebook and a common time frame to sporadically transmit their signals to a base station (BS) \cite{polyanskiy2017perspective,Ahmadi_PHD_thesis}. Under this framework, the BS cares only about the transmitted messages, i.e., the identity of users is not important, and the per-user probability of error (PUPE) is used as the performance criterion. Many coding schemes have been devised for URA over a Gaussian multiple-access channel (GMAC)~\cite{Ozates2024ODMA,pradhan2020polar,ahmadi2021random}, and block-fading channels~\cite{Ahmadi2023Unsourced, ahmadi2021Unsourced,Ozates2022slotted,Ahmadi2023RIS,Ahmadi2024Integrated}.

Reconfigurable intelligent surface (RIS) is a promising technology developed for providing high spectral efficiency and energy savings for 5G and beyond wireless communication systems. Specifically, a passive RIS equipped with many low-cost passive elements, which can intelligently tune the phase shift of the incident electromagnetic waves, and reflect them in a desired direction without any amplification, can improve the efficiency of the network by enabling line-of-sight paths between the transmitters and the receivers in problematic environments with many blocking obstacles~\cite{Laue2021IRS,Shao2022reconf,Yan2020Passive,Shen2019Secrecy,Swindlehurst2022Channel,Huang2019Reconfigurable,Wu2019Intelligent,Wu2018Intelligent}. 

URA schemes in the literature consider direct links between all the users and the BS \cite{Ozates2024ODMA,pradhan2020polar,ahmadi2021random,Ahmadi2023Unsourced, ahmadi2021Unsourced,Ozates2022slotted,Ahmadi2023RIS}; however, in certain environments, the direct link between some users and the BS may be blocked or significantly attenuated. Therefore, the use of RIS can improve user connectivity in the URA by creating high-quality links between the BS and the users.

One of the most crucial problems in RIS-aided communication systems is the design of RIS reflecting coefficients to achieve the best system performance. This is solved via different approaches such as alternating optimization \cite{Yan2020Passive,Shen2019Secrecy}, Gaussian randomization \cite{Laue2021IRS,Shao2022reconf}, gradient descent \cite{Huang2019Reconfigurable}, and semidefinite relaxation (SDR) \cite{Wu2019Intelligent,Wu2018Intelligent}. To design the RIS elements, a reliable knowledge of channel state information (CSI) is required. For this purpose, many grant-based multiple-access schemes solve the channel estimation problem by providing each user with an individual time slot~\cite{Wei2021Channel,Swindlehurst2022Channel, Hu2021Two,Ruan2022Approximate, Guo2022Uplink, Shi2022Triple}. Specifically, they assign each user an orthogonal pilot so that its CSI can be estimated without suffering from interference due to the simultaneous transmission of other users. However, since users cannot be assigned dedicated pilot sequences, orthogonal transmission is not feasible for URA, hence new solutions are needed.

Authors in \cite{Shao2022reconf} propose a RIS-aided URA scheme, where a tensor-based model is employed. Specifically, every user transmits a rank-1 tensor, and as a result, the received signal is a tensor with a rank equal to the number of active users, perturbed by AWGN noise. Utilizing a coupled tensor decomposition technique at the receiver, the signals from distinct users along with their respective channel coefficient vectors are jointly estimated. In addition to channel estimation and data detection, they also propose a RIS design algorithm. Despite a suitable performance in the case of Rayleigh RIS-BS channel with a full-rank channel coefficient matrix, the approach encounters a notable performance deterioration in cases when the RIS-BS channel matrices are rank-deficient.

In this paper, we propose a slotted URA scheme facilitated by a passive RIS coupled with the necessary channel estimation, RIS design, and pilot and data detection algorithms. In the proposed RIS-aided unsourced multiple access scheme (RISUMA), every user transmits a signal consisting of a pilot, appended to a polar codeword. The decoding process at the receiver takes place in two phases: the RIS configuration phase and the data phase. The RIS configuration phase is responsible for jointly detecting the active pilots and estimating the CSI via the newly proposed joint pilot detection and channel estimation (JDCE) algorithm, as well as designing the reflection coefficients of the RIS elements. During the data phase, the actual data transmission takes place. As stated above, unlike the grant-based schemes, where each identified user's CSI is estimated without any interference from the other users, in the URA, pilots from multiple users interfere with each other due to the unsourced nature of transmission for which there is no cooperation between the users and the BS, or among the users themselves. Therefore, the CSI estimation task becomes more challenging. Nevertheless, the proposed JDCE algorithm is able to perform channel estimation successfully without any prior knowledge of the number of active users or their transmitted pilots. \\
\indent In the proposed RIS-aided URA set-up, we also need to design the RIS reflection coefficients, which is not an issue in the standard URA. In particular, we propose two RIS design algorithms: alternating SDR (ASDR) and adaptive eigenvalue decomposition (AEVD). In the former, we alternatively solve a standard SDR problem, and update the RIS coefficients at each iteration. For the latter, we employ an eigenvalue decomposition to find the appropriate RIS coefficients of each active user at each iteration, and then combine the resulting coefficients of different users. We show that the AEVD algorithm performs similarly to the ASDR, while having a lower computational complexity. On the other hand, for the encoder and decoder designs, similar solutions with other URA schemes can be adopted~\cite{ahmadi2021Unsourced,Ahmadi2023Unsourced}. Specifically, in this paper, we employ polar codes along with successive interference cancellation (SIC) for the data phase to recover the transmitted messages. The aforementioned algorithms are devised for the situation where the direct communication links between the user and the BS are completely blocked. Additionally, we establish an approximate achievability bound for the RIS-based URA scheme, serving as a benchmark. We also extend the proposed algorithms to the scenario with direct user-BS paths. We demonstrate that the newly proposed RISUMA method surpasses the CTAD algorithm introduced in \cite{Shao2022reconf}, which currently stands as the state-of-the-art within the RIS-assisted URA schemes.

\indent Our contributions are as follows:
\begin{itemize}
\item  We propose a RIS-assisted URA scheme, dubbed as RISUMA, including appropriate encoding/decoding blocks, considering the direct user-BS link to be completely blocked. The proposed scheme offers superior performance compared to the state-of-the-art.

\item A joint pilot detector and channel estimator algorithm (called JDCE) is proposed, which detects the active pilots and estimates the corresponding users' channel coefficients without the knowledge on the number of active users or their identities. The proposed JDCE also demonstrates the ability to reduce interference, consequently exhibiting favorable detection and channel estimation performance within the URA framework.
\item Two RIS phase shift design algorithms (namely, ASDR and AEVD), with the objective of increasing the signal-to-interference-plus-noise ratio (SINR) of the input to the polar decoder, are described.

%\item We also decode bit sequences of active users via a single user polar list decoder, while treating other user's signal as noise. Then, an appropriate  SIC is adopted to decrease the interference level of the system by removing the contribution of the successfully decoded messages from the received signal.

%\item To study the proposed scheme's performance, we analyse the computational complexity of RISUMA.
\item The proposed solution is also extended to the case where a direct link also exists between the users and the BS. In the resulting scheme, the direct user-BS channels as well as the cascaded channels are estimated. Then, the RIS design algorithm and the decoder are built based on the overall channels of the users.  
\item An information theoretic achievability bound for RIS-based URA, considering a scenario with a completely blocked user-BS path, is established.
\end{itemize}

The paper is organized as follows. In Section \ref{SystemModel}, we describe the system model. Section \ref{Sec53} presents the proposed RIS-aided URA scheme for the case with blocked user to BS channels. Specifically, we describe the proposed encoder, pilot detector, channel estimator, RIS design, decoding procedures, and achievability bound in detail. In Section \ref{Sec54}, we extend the scheme in Section \ref{Sec53} to the case with additional user-BS links. In Section \ref{sectionSimulationResults}, we present extensive simulation results, and conclude the paper in Section \ref{Conclustions}.

\textit{Notation:} Lower-case and upper-case boldface letters are used to denote a vector and a matrix, respectively; $\mathrm{diag}(\mathbf{t})$ and $\mathbf{I}_N$ represent a diagonal matrix with elements of vector $\mathbf{t}$ in its diagonal, and an $N\times N$ identity matrix, respectively; $\mathrm{Re}(\cdot)$ and $ \mathrm{trace}(\cdot)$ denote the real part and trace of a matrix, respectively; $\mathbf{T}^T$, $\mathbf{T}^H$, and $\mathbf{T}^*$ are the transpose, the Hermitian, and the conjugate of matrix $\mathbf{T}$, respectively; $\left[ \mathbf{T}\right]_{(i,j)}$ refers to the element in the $i$th row and the $j$th column of $\mathbf{T}$; $\left[ \mathbf{T}\right]_{(i,:)}$ and $\left[ \mathbf{T}\right]_{(:,j)}$ denote the elements in the $i$th row and $j$th column of $\mathbf{T}$, respectively; $\left[ \mathbf{t}\right]_{i}$ represents the $i$th element of the vector $\mathbf{t}$; $\mathrm{vec}(\cdot)$ is the vectorization operator; $\otimes$ denotes the Kronecker product; $|\mathcal{A}|$ denotes the cardinality of the set $\mathcal{A}$;  $|\mathbf{A}|$ denotes the determinant of the matrix $\mathbf{A}$; $\mathcal{CN}(\mathbf{0},\mathbf{B})$ denotes the zero-mean circularly symmetric complex Gaussian random variable with covariance matrix $\mathbf{B}$; $U(c,d)$ is the continuous uniform distribution on the internal $[c,d]$; and, the set $\mathcal{T}(N) = \{-1,-1+\Delta_N,-1+2\Delta_N,-1+3\Delta_N,...,-1+(N-1)\Delta_N\}$ with $\Delta_N = 2/N$.

\begin{figure}[t!]%[h!]
	\centering
	\includegraphics[width=1\linewidth]{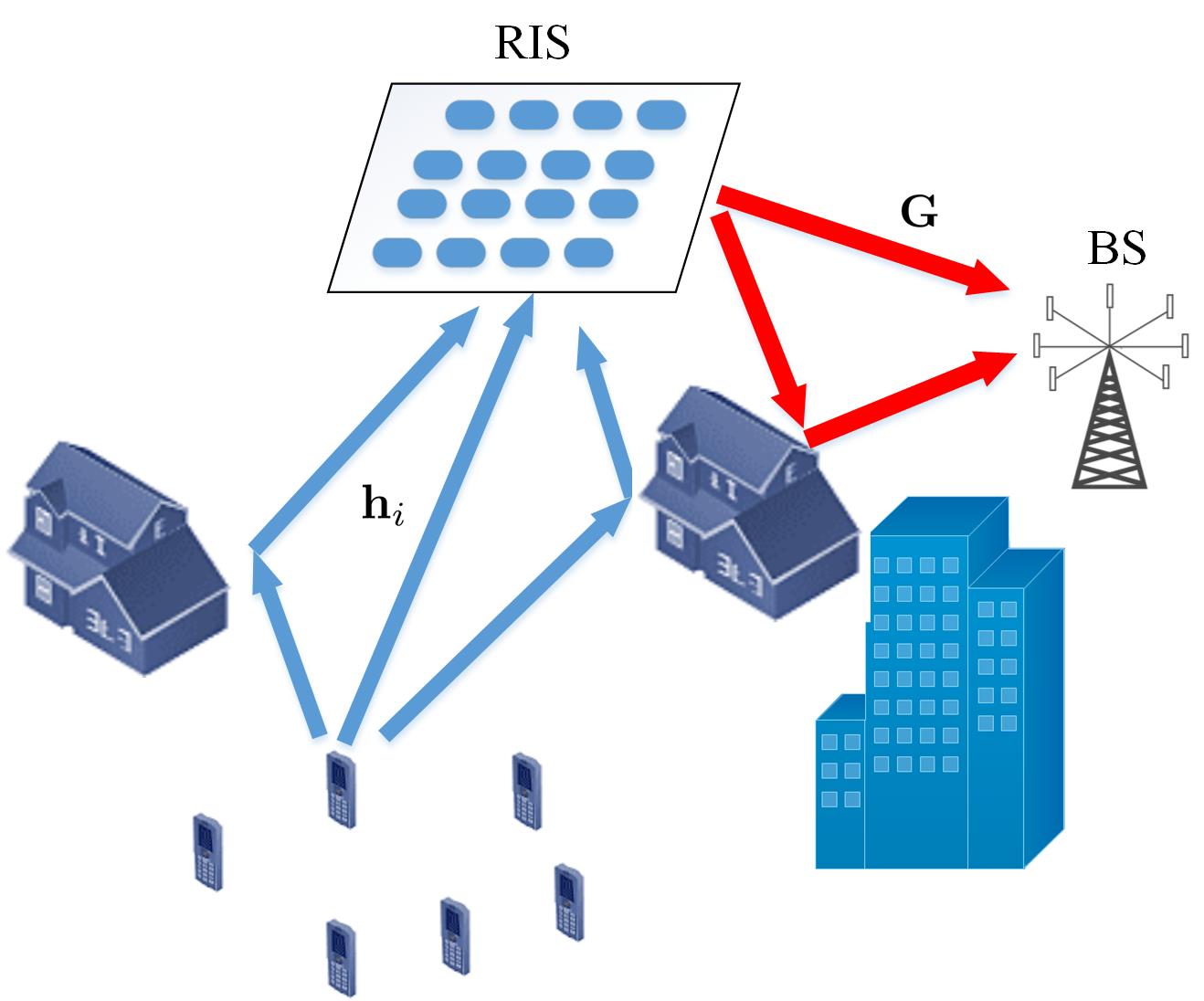}
	\caption{{ Illustration of a RIS-aided URA system.}	}
	\label{Fig_subplot}
\end{figure}
\section{System Model}
\label{SystemModel}
Consider a RIS-aided URA system as depicted in Fig. \ref{Fig_subplot}, where a BS with $M$ receive antennas serves $K_T$ single-antenna users (of which only $K_a$ are active in any transmission frame) with the help of a passive $N$-element RIS. We assume that both the BS and the RIS are equipped with a uniform planer array (UPA) \cite{Wei2021Channel}. Active users send $B$ bits of information to the BS through $n$ channel uses. Employing the Saleh-Valenzuela channel model, the RIS-BS channel is written as \cite{Shao2022reconf}
\begin{align}
\mathbf{G} =\sqrt{MN}\sum_{l =1}^{ L_G }\mu_{l}\mathbf{a}_{M}(\phi_{r,l},\psi_{r,l})^T\mathbf{a}_N(\phi_{t,l},\psi_{t,l}) \in \mathbb{C}^{M\times N},
\label{eqs1}
\end{align}
where $(\phi_{r,l},\psi_{r,l})$ are the azimuth and elevation angles of arrival (AOA) at the BS from the $l$th path, $(\phi_{t,l},\psi_{t,l})$ are the azimuth and elevation angles of departure (AOD) from the RIS to the BS through the $l$th path, $L_G$ is the number of paths from the RIS to the BS, $\mu_{l}$ denotes the $l$th path's gain which is modeled as a circularly symmetric complex Gaussian random variable, i.e., $\mu_l \sim \mathcal{CN}(0, L_0d_l^{-\alpha_{\mathrm{PL}}})$~\cite{Wu2020Beamforming}, $d_l$ is the length of the $l$th path, $L_0$ and $\alpha_{\mathrm{PL}}$ denote the path loss at the reference distance and the path loss exponent, respectively, and $\mathbf{a}_N(\phi,\psi)$ is the array steering vector of an $N_1\times N_2$ UPA ($N=N_1N_2$) which is represented as \footnote{For the vector $\mathbf{t}$, we denote the element-wise exponentiation by $e^{\mathbf{t}}$.}
\begin{align}
    \mathbf{a}_N(\phi,\psi) =\bar{\mathbf{a}}_{N,\bar{\phi},\bar{\psi}}= \dfrac{1}{\sqrt{N}}e^{-j2\pi \bar{\phi} \mathbf{n}_1}\otimes e^{-j2\pi \bar{\psi} \mathbf{n}_2}\in \mathbb{C}^{1\times N},\label{eq_steer}
\end{align}
where $\mathbf{n}_1 = \dfrac{d}{\lambda}[0,...,N_1-1]$, $\mathbf{n}_2 = \dfrac{d}{\lambda}[0,...,N_2-1]$, $\bar{\phi}=\sin{(\phi)} \cos{(\psi)}$, $\bar{\psi}=\sin{(\psi)}$, $\lambda$ denotes the carrier wavelength, and $d$ is the antenna spacing. Note also that $\mathbf{a}_M(\phi,\psi)$ is the array steering vector of an $M_1 \times M_2$ UPA ($M=M_1M_2$), obtained by replacing $N$ by $M$ in \eqref{eq_steer}. Similarly, the channel from the $i$th user to the RIS can be expressed as
\begin{align}
\mathbf{h}_i = \sqrt{N} \sum_{f_i =1}^{ L_{R,i} }\mu_{f_i}\mathbf{a}_{N}(\phi_{i,f_i},\psi_{i,f_i}) \in \mathbb{C}^{1\times N},
\label{eqs3}
\end{align}
where $L_{R,i}$ is the number of paths between the $i$th user and the RIS, $(\phi_{i,f_i},\psi_{i,f_i})$ are the azimuth and elevation AOAs at the RIS for the $f_i$th path of the $i$th user's signal, with the distance $d_{f_i}$ and the corresponding path gain $\mu_{f_i}$. The path-loss model is the same as the one for the RIS-BS channel. The channel between the RIS and the BS is assumed to be perfectly known. This can be justified as the RIS and the BS are stationary, so the channel between them changes very slowly, and it can be well estimated with negligible overhead.

\begin{figure}[t!]%[h!]
	\centering
	\includegraphics[width=1\linewidth]{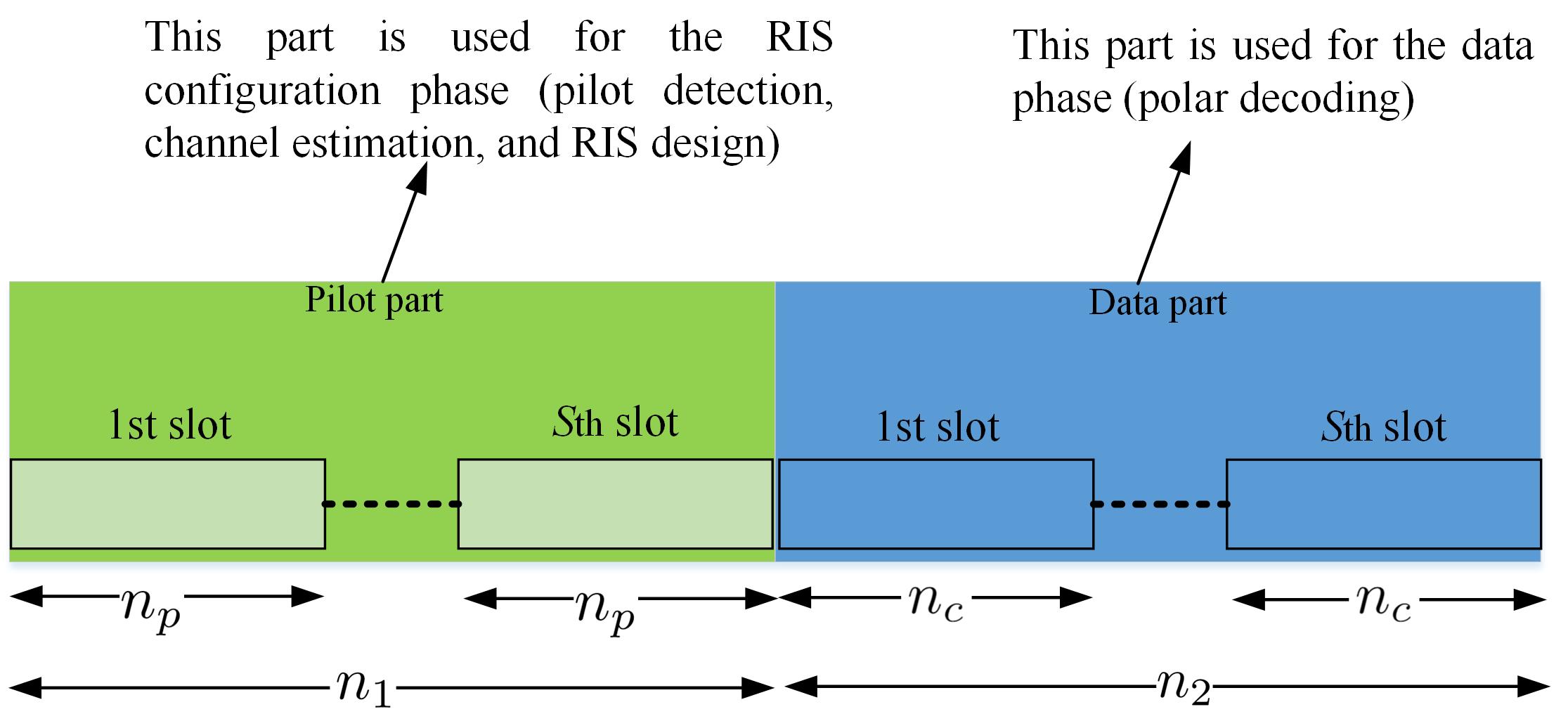}
	\caption{{\small Transmission structure of the proposed URA scheme.}	}
	\label{fig_TRANSMISSION}
\end{figure}
\section{RISUMA with Blocked User-BS links}
\label{Sec53}
Let $\mathbf{y}_t \in \mathbb{C}^{M\times 1}$ denote the uplink signal received at the BS at time $t$. Assuming that the direct link between the users and the BS is completely blocked, and the channel coefficients are constant throughout the frame, the received signal can be written as \cite{Wei2021Channel}
\begin{align}
    \mathbf{y}_t = \sum_{i=1}^{K_a} \mathbf{G}\mathrm{diag}(\mathbf{h}_i)\mathbf{w}_t x_{i,t}+ \mathbf{z}_t, \in \mathbb{C}^{M\times 1} , \ t=1,...,n, \label{cascadedChannel}
\end{align}
where $\mathbf{w}_t\in \mathbb{C}^{N\times 1}$ denotes the reflection coefficient vector of the RIS at time $t$ with $\left|[\mathbf{w}_t]_{i}\right|=1$ (passive RIS assumption), $x_{i,t}\in \mathbb{C}$ is the symbol transmitted by the $i$th user at time $t$, and $\mathbf{z}_t\sim \mathcal{CN}(\mathbf{0}, \sigma_z^2\mathbf{I}_M)$ represents the noise vector at time $t$.

The proposed coding scheme divides the entire frame into slot pairs. Each active user randomly selects only one slot pair to transmit its pilot and data signals as illustrated in Fig. \ref{fig_TRANSMISSION}. At the receiver side, transmitted messages are decoded in two phases: the RIS configuration phase and the data phase. In the RIS configuration phase, active pilots in each slot are identified, and the corresponding CSI between the users and the RIS is estimated. Given the estimated channel of each active user, the BS designs the RIS reflection coefficients, and transmits them back to the RIS unit. After the phase shifts of the RIS elements are adjusted accordingly, the active users transmit their encoded signals in the data transmission phase. The assumption is that the channels are quasi static, i.e., they do not change over a frame (including the pilot and data transmission parts).

In the following, we describe the proposed encoder and decoder structures in detail.
\subsection{Encoder}
\label{encoder}
As shown in Fig. \ref{fig_TRANSMISSION}, in the proposed transmission scheme, a length-$n$ frame is divided into pilot and data parts with lengths $n_1$ and $n_2$, respectively, where $n_2+n_1 = n$. We also divide each part into $S$ slots of lengths $n_{p}= n_1/S$ and $n_{c}= n_2/S$. The $i$th user divides its $B$ bits of information into two parts: pilot part $\omega_{p_i}\in \{0,1\}^{1\times B_p}$ and data part $\omega_{c_i}\in \{0,1\}^{1\times B_c}$, where $B = B_p+B_c$. Every user obtains a preamble $\mathbf{b}_i\in \mathbb{C}^{1\times n_s}$ and a pilot $\mathbf{p}_i\in \mathbb{C}^{1\times n_p}$ by mapping $\omega_{p_i}$ to the rows of the preamble codebook $\mathbf{B}\in \mathbb{C}^{2^{B_p}\times n_s}$ and the pilot codebook $\mathbf{P}\in \mathbb{C}^{2^{B_p}\times n_p}$, respectively. The $i$th user adds cyclic redundancy check (CRC) bits to $\omega_{c_i}$, and encodes the result with an $\left(n_d, \  B_c +r\right)$ polar code, where $r$ is the number of CRC bits. The encoded bits are then modulated using binary phase shift keying (BPSK), which results in $\mathbf{v}_i=[v_{i,1},v_{i,2},...v_{i,n_d}]\in \{\pm1\}^{1\times n_d}$. To construct the signal of the data part, the encoder spreads each symbol of the modulated signal by its preamble \cite{pradhan2020polar,ahmadi2021random}, i.e., $\mathbf{c}_i  = \mathbf{v}_i \otimes \mathbf{b}_i\in \mathbb{C}^{1\times n_c }$, where $n_c = n_sn_d$. The $i$th user randomly selects a slot index (denoted by $\zeta_i$), and transmits $\mathbf{p}_i$ and $\mathbf{c}_i$ through the $\zeta_i$th slot of the pilot and data parts, respectively. 

Using the signal model in \eqref{cascadedChannel}, the received signal at the $s$th slot of the pilot part can be written as
	\begin{align}
 \label{receivedP}
	    \mathbf{Y}_{p} &= \sqrt{ P_p}\sum_{i\in\mathcal{S}_s}^{} \mathbf{G}\mathrm{diag}(\mathbf{h}_i )\mathbf{W}_{p_s}\mathrm{diag}(\mathbf{p}_i)+ \mathbf{Z}_{p}\in \mathbb{C}^{M\times n_p},
	\end{align}
 where $P_p$ denotes the signal power in the pilot part, $\mathcal{S}_s$ is the set of users in the $s$th slot, $\mathbf{W}_{p_s} = [\mathbf{w}_{j_{1,s}}, ...,\mathbf{w}_{j_{n_p,s}}]$ contains the RIS reflecting coefficients in the $s$th slot of the pilot part, $\mathbf{Z}_{p}=[\mathbf{z}_{j_{1,s}},...,\mathbf{z}_{j_{n_p,s}}]$, and $j_{t,s} = (s-1)n_p+t$. We omit the slot index from matrices for notational simplicity. Similarly, the received signal at the $s$th slot of the data part corresponding to the $f$th ($f  = 1,2,...,n_c$) symbol of the modulated signal can be written as
 \begin{align}
 \nonumber
	    \mathbf{Y}_{c,f} = \sqrt{ P_c}\sum_{i\in\mathcal{S}_s}^{} \mathbf{G}\mathrm{diag}(\mathbf{h}_i )\mathbf{W}_{c_s}&\mathrm{diag}(\mathbf{b}_i) v_{i,f}\\
     &+ \mathbf{Z}_{c,f}\in \mathbb{C}^{M\times n_s},  \label{receivedc}
\end{align}
 where $P_c$ denotes the signal power in the data part, $\mathbf{W}_{c_s}\in \mathbb{C}^{N\times n_s}$ contains the RIS reflecting coefficients in the $s$th slot of the data part, $\mathbf{Z}_{c,f}=[\mathbf{z}_{k_{1,s}},...,\mathbf{z}_{k_{n_s,s}}]$, and $k_{t,s} = Sn_p+(s-1)n_c+(f-1)n_s+t$. Note that we consider two choices for selecting $\mathbf{W}_{c_s}$: 1) $\mathcal{C}_0$: the RIS coefficient vectors remain constant during each symbol's duration, 2) and $\mathcal{C}_1$: the RIS coefficient vectors vary symbol by symbol, i.e.,

\begin{align}
 \mathbf{W}_{c_s}=    \left\{\begin{matrix}
 [\mathbf{w}_{f,s}, ...,\mathbf{w}_{f,s}]   \ \ \  &\text{ for } \mathcal{C}_0\\ 
  [\mathbf{w}_{k_{1,s}}, ...,\mathbf{w}_{k_{n_s,s}}]& \text{ for } \mathcal{C}_1
\end{matrix}\right.. \label{W_c_structures}
\end{align}
 \subsection{Decoder} 
 \label{decoder_Blocked_UB}
\subsubsection{Joint Pilot Detection and Channel Estimation (JDCE)}
\label{Sec_ch_estimation}
In the following, we describe the steps of the JDCE algorithm. It is clear from \eqref{eqs3} that detecting all the active paths (finding angles of active paths), and estimating the corresponding path gains give us an estimate of the user-RIS channel $\mathbf{h}_i$. The pilot with the highest energy is detected first. Then, we identify the path with the highest energy over which the detected pilot arrives at the RIS. We also estimate the gain corresponding to the detected path using the least-squares (LS) technique. Finally, we remove the contribution of the detected pilot-path pair from the received signal, and repeat the procedure on the remaining signal to detect the next pilot-path pair. Note that since the residual received signal is updated at each iteration, we denote it by $\mathbf{Y}_{p}^j$, where $j$ is the iteration index. Details of the proposed JDCE algorithm are given below:
\begin{itemize}
\item \textbf{Step 1 }(pilot detection): the pilot with the highest probability of existence is detected using the energy detection approach as in \cite{pradhan2020polar,ahmadi2021random}
\begin{align}
    \hat{k}=\argmax_{k\in{1,2,...,2^{B_p}}} \dfrac{\mathrm{trace}(\mathbf{Q}_k\mathbf{R}_{pj} \mathbf{Q}_k^H)}{\mathrm{trace}(\mathbf{Q}_k\mathbf{Q}_k^H)}, \label{energyDetector}
\end{align}
where $\mathbf{R}_{pj} ={\mathbf{Y}_{p}^j}^H\mathbf{Y}_{p}^j $, $\mathbf{Q}_k=\mathbf{W}_{p_s}\mathrm{diag}(\bar{\mathbf{p}}_k)$, and $\bar{\mathbf{p}}_k$ is the $k$th row of the codebook $\mathbf{P}$.
 \item \textbf{Step 2 }(path detection): By solving the following problem, we find the strongest path through which the $\hat{k}$th pilot (the detected pilot) arrives at the RIS
\begin{align}
    (\hat{l},\hat{q})=\argmax_{l\in \mathcal{T}(N_1),q\in \mathcal{T}(N_2)} \dfrac{\mathrm{trace}(\mathbf{F}_{l,q}\mathbf{R}_{pj} \mathbf{F}_{l,q}^H)}{\mathrm{trace}(\mathbf{F}_{l,q}\mathbf{F}_{l,q}^H)}, \label{pathDetect}
\end{align}
where $\mathbf{F}_{l,q}= \mathbf{G}\mathrm{diag}(\bar{\mathbf{a}}_{N,l,q} )\mathbf{W}_{p_s}\mathrm{diag}(\bar{\mathbf{p}}_{\hat{k}})$, and $\bar{\mathbf{a}}_{N,l,q}$ is defined in \eqref{eq_steer}.
\item \textbf{Step 3 }(SIC): from \eqref{receivedP}, we can infer that the contribution of the received signal corresponding to the detected pilot-path pair in the $j$th iteration can be obtained as
\begin{align}
    \mathbf{U}_j =  \sqrt{P_p}\mathbf{G}\mathrm{diag}(\bar{\mathbf{a}}_{N,\hat{l},\hat{q}} )\mathbf{W}_{p_s}\mathrm{diag}(\bar{\mathbf{p}}_{\hat{k}})\in \mathbb{C}^{M\times n_p}.
\end{align}
Vectorizing both sides of \eqref{receivedP}, the received signal vector at the initial iteration ($\mathbf{y}_{p}^1=\mathrm{vec}(\mathbf{Y}_{p}^1) \in \mathbb{C}^{Mn_p\times 1}$) can be written as
\begin{align}
\mathbf{y}_{p}^1 &= \sum_{{j^\prime} =1}^{j} \bar{\mu}_{j^\prime} \mathbf{u}_{j^\prime}+ \mathbf{z}_{p}^{\prime},
	\end{align}
where $\mathbf{u}_j =  \mathrm{vec}(\mathbf{U}_j) \in \mathbb{C}^{Mn_p\times 1}$, $\bar{\mu}_{j^{\prime}} \in \mathbb{C}$ is the path gain corresponding to the pilot-path pair detected at the $j^{\prime}$th iteration, and $\mathbf{z}_{p}^{\prime}\in \mathbb{C}^{Mn_p\times 1}$ contains the signals corresponding to the pilot-path pairs that are not detected yet along with the vectorized noise term. The path gains are estimated using LS as
\begin{align}
    \mathbf{\hat{m}} =  (\mathbf{\bar{U}}^H\mathbf{\bar{U}})^{-1}\mathbf{\bar{U}}^H\mathbf{y}_{p}^1,\label{pathGainEst}
\end{align}
where the $k$th column of $\mathbf{\bar{U}}\in \mathbb{C}^{Mn_p\times j}$ is $\mathbf{u}_k$. Then, the contribution of $ \mathbf{\bar{U}}$ is removed from the initially received signal to obtain
\begin{align}
        \mathbf{y}_{p}^{(j+1)} = \mathbf{y}_{p}^1-       \mathbf{\bar{U}}\mathbf{\hat{m}}.\label{SIC_blocked}
\end{align}
Finally, $\mathbf{y}_{p}^{(j+1)}$ is converted back to an $M\times n_p$ matrix $\mathbf{Y}_{p}^{(j+1)}$ before being passed to Step~1 for the $(j+1)$th iteration. The procedure is stopped if the condition 
\begin{align}
\dfrac{1}{Mn_p}\dfrac{\|\mathbf{y}_{p}^{(j+1)}\|^2-\|\mathbf{y}_{p}^{j}\|^2}{\sigma_z^2}\leq \alpha_1 \label{StoppingDetector}
\end{align}
is satisfied, where $\alpha_1\in \mathbb{R}^+$ is a threshold.
\end{itemize}
\subsubsection{Data Phase}
\label{sec_polar}
We now devise an iterative algorithm to detect the message bits $\omega_{c_i}$ using the received signal corresponding to the data part. The algorithm 1) generates log-likelihood ratios (LLRs) of the polar coded bits using the estimated channel coefficients, 2) passes the LLRs to the polar list decoder, 3) removes the contribution of the successfully decoded users (users whose detected messages satisfy the CRC check) from the received signal. The remaining received signal is passed back to the first step to generate a new LLR. Details of the proposed algorithm are given below.\\
We can rewrite the received signal in \eqref{receivedc} as
\begin{align}
    	    \mathbf{Y}_{c,f} = \sum_{i\in\mathcal{S}_s}^{}\mathbf{T}_iv_{i,f}+ \mathbf{Z}_{c,f}, \label{received_approx}
\end{align}
where $\mathbf{T}_i  =  \sqrt{ P_c}\mathbf{G}\mathrm{diag}(\mathbf{h}_i )\mathbf{W}_{c_s} \mathrm{diag}(\mathbf{b}_i)\in \mathbb{C}^{M\times n_s}$. Vectorizing both sides of \eqref{received_approx}, we write
	\begin{align}
 \nonumber
	    \mathbf{y}_{c,f} &= \sum_{i\in\mathcal{S}_s}^{}  \mathbf{t}_iv_{i,f}+ \mathbf{z}_{c,f}%\label{vec_received11} 
     \\
     &=   \mathbf{\Bar{T}}\mathbf{\tilde{v}}_f+ \mathbf{z}_{c,f},\label{vec_received}
	\end{align}
 where 
 \begin{align}
 \nonumber
     \mathbf{t}_i&=\mathrm{vec}(\mathbf{T}_i)\\
     &=\sqrt{ P_c}\mathrm{vec}(\mathbf{G}\mathrm{diag}(\mathbf{h}_i )\mathbf{W}_{c_s} \mathrm{diag}(\mathbf{b}_i))\in \mathbb{C}^{Mn_s\times 1},\label{t_i_showing}
 \end{align}
where $\mathbf{y}_{c,f}=\mathrm{vec}(\mathbf{Y}_{c,f})\in \mathbb{C}^{Mn_s\times 1}$, $\mathbf{z}_{c,f}=\mathrm{vec}(\mathbf{Z}_{c,f})\in \mathbb{C}^{Mn_s\times 1}$, $v_{i,f}$ is the $i$th row of $\mathbf{\tilde{v}}_f\in \mathbb{C}^{|\mathcal{S}_s|\times 1}$, and $\mathbf{t}_i$ is the $i$th column of $\mathbf{\bar{T}}\in \mathbb{C}^{Mn_s\times|\mathcal{S}_s|}$. We estimate $\mathbf{\tilde{v}}_f$ using the minimum mean square error (MMSE) criterion as
\begin{align}
   \mathbf{\hat{v}}_f= [ \hat{v}_{1,f},\hat{v}_{2,f},...,\hat{v}_{|\mathcal{S}_s|,f}]^T=  \mathbf{\Bar{T}}^H\mathbf{\hat{R}}^{-1}\mathbf{y}_{c,f}, \label{MMSE}
\end{align}
where 
\begin{align}
\mathbf{\hat{R}}=\mathbb{E}\{\mathbf{y}_{c,f}\mathbf{y}_{c,f}^H\}= \mathbf{\Bar{T}} \mathbf{\Bar{T}}^H+\sigma_z^2\mathbf{I}_{Mn_s}.\label{R_estimate}    
\end{align}
In a similar way as in \cite{pradhan2020polar}, the LLR of the $f$th symbol of the $i$th user is approximated as 
\begin{align}
    \hat{g}_{i,f} \approx 2\mathrm{Re}\left(\hat{v}_{i,f}/\delta_i\right),
\end{align}
where $\delta_i$ is the $i$th diagonal element of the matrix $\mathbf{\Sigma} =
\mathbf{I}_{|\mathcal{S}_s|}- \mathbf{\Bar{T}}^H\mathbf{\hat{R}}^{-1}\mathbf{\Bar{T}}$. The obtained LLR values are then passed to the polar list decoder. Finally, the contributions of users whose decoded message sequences satisfy the CRC check are removed from the received signal employing
\begin{align}
    	    \mathbf{Y}_{c,f} =\mathbf{Y}_{c,f}- \mathbf{T}_i\bar{v}_{i,f}, \label{SIC}
\end{align}
where $\bar{v}_{i,f}$ is the $f$th symbol of the $i$th user obtained by encoding and modulating its successfully decoded message. The remaining signal $ \mathbf{Y}_{c,f}$
is passed back to the MMSE estimator in \eqref{MMSE} for the next iteration. The step terminates if no user is successfully decoded during an iteration.

\subsubsection{RIS Design}
\label{SectionRISdesign}
Unlike the RIS phase shift coefficient matrix $\mathbf{W}_{p_s}$, which is arbitrarily selected, in this section, we propose algorithms for designing the RIS phase shift matrix for use in the data part, denoted by $\mathbf{W}_{c_s}$. Note that since the RIS algorithm is designed based on the decoder, we introduce it here, after studying the data phase of the decoder. However, in practice, the RIS coefficients are obtained before the actual data transmission starts. 

We know from Section \ref{sec_polar} that the MMSE estimate of the BPSK signal is fed to the polar decoder. Thus, improving the SINR at the output of the MMSE block is a good way to decrease the decoding error of each user. Plugging \eqref{vec_received} into \eqref{MMSE}, we can obtain the MMSE estimate of the $f$th symbol of the $i$th user's codeword as
\begin{align}
     \hat{v}_{i,f}&=    \mathbf{t}_i^H   \mathbf{\hat{R}}^{-1}\mathbf{t}_iv_{i,f}+\mathbf{t}_i^H   \mathbf{\hat{R}}^{-1}\left(\sum_{k\in\mathcal{S}_s, k\neq i}^{} \mathbf{t}_kv_{k,f}+ \mathbf{z}_{c,f}\right),
\end{align}
where the first and second terms are the signal and interference-plus-noise terms, respectively, whose powers can be computed as
\begin{align}
\nonumber
\sigma_{s,i}^2    & = \mathbb{E}\{\|\mathbf{t}_i^H   \mathbf{\hat{R}}^{-1}\mathbf{t}_iv_{i,f}\|^2\}\\
    &=( \mathbf{t}_i^H   \mathbf{\hat{R}}^{-1}\mathbf{t}_i)^2, \label{20}\\\nonumber
\end{align}
    \begin{align}
    \sigma_{\mathrm{IN},i}^2 & = \mathbb{E}\{\|\hat{v}_{i,f}\|^2\}-\sigma_{s,i}^2\\\nonumber
    & =  \mathbb{E}\{\|\mathbf{t}_i^H   \mathbf{\hat{R}}^{-1}\mathbf{y}_{c,f}\|^2\}-\sigma_{s,i}^2\\\nonumber
    & =  \mathbf{t}_i^H   \mathbf{\hat{R}}^{-1}\mathbb{E}\{\mathbf{y}_{c,f}   \mathbf{y}_{c,f}^H \} \mathbf{\hat{R}}^{-1}\mathbf{t}_i-\sigma_{s,i}^2\\\nonumber
    & =  \mathbf{t}_i^H   \mathbf{\hat{R}}^{-1}\mathbf{t}_i-\sigma_{s,i}^2\\
    & =  \sigma_{s,i}-\sigma_{s,i}^2.
\end{align}
Therefore, the SINR of the $i$th user at the output of the MMSE estimator (input to the polar decoder) can be calculated as $\beta_i = \sigma_{s,i}/(1-\sigma_{s,i})$. Using this SINR term, the decoding error probability of the $i$th user is approximated as \cite{Polyanskiy2010Channel}
\begin{align}
 e_{r,i}=    \mathcal{F}(\sigma_{s,i}),\label{er_i}
\end{align}
where
\begin{align}
 \mathcal{F}(x)= Q\left(\dfrac{ 0.5\log_2\left(1+x^\prime  \right)-\dfrac{B_c +r}{n_d}}{ \sqrt{ \dfrac{1}{n_d}\dfrac{x^\prime (x^\prime +2)\log_2^2e}{2(x^\prime +1)^2}}}\right)  , \label{Fx}
\end{align}
with $x^\prime = x/(1-x)$, and $Q(.)$ denotes the standard $Q$-function. Using \textit{vec trick} property~\cite{Roth1934On}, $\mathrm{vec}(\mathbf{A}_1\mathbf{A}_2\mathbf{A}_3)=(\mathbf{A}_3^T\otimes \mathbf{A}_1)\mathrm{vec}(\mathbf{A}_2)$, we can write $\mathbf{t}_i$ in \eqref{t_i_showing} as 
\begin{align}
    \mathbf{t}_i  =   \sqrt{ P_c}\mathbf{E}_i\mathbf{w}_c,
\end{align}
where $\mathbf{E}_i$ and $\mathbf{w}_c$ are obtained as
\begin{align}
\mathbf{E}_i=    \left\{\begin{matrix}
   \left(\mathbf{b}_i^T\otimes \mathbf{G}\mathrm{diag}(\mathbf{h}_i )\right)\in \mathbb{C}^{Mn_s\times N} \ \ \ \ \ \ \  \ & \text{ for } \mathcal{C}_0\\ 
 \left(\mathrm{diag}(\mathbf{b}_i)\otimes \mathbf{G}\mathrm{diag}(\mathbf{h}_i )\right)\in \mathbb{C}^{Mn_s\times Nn_s}& \text{ for }\mathcal{C}_1
\end{matrix}\right., \label{W_c_structures2}
\end{align}
and
\begin{align}
 \mathbf{w}_c=    \left\{\begin{matrix}
   [\mathbf{W}_{c_s}]_{(:,i)}\in  \mathbb{C}^{ N\times 1}  \ \ \  & \text{ for } \mathcal{C}_0\\ 
 \mathrm{vec}(\mathbf{W}_{c_s})\in  \mathbb{C}^{ Nn_s\times 1} & \text{ for }\mathcal{C}_1
\end{matrix}\right., \label{W_c_structures3}
\end{align}
respectively. Hence
\begin{align}
\sigma_{s,i}  =  \mathbf{w}_c^H\mathbf{C}_i(\mathbf{w}_c)\mathbf{w}_c, \label{vec_ti}
\end{align}
 where 
 \begin{align}
     \mathbf{C}_i(\mathbf{w}_c)=\mathbf{E}_i^H \mathbf{\hat{R}}^{-1}\mathbf{E}_i.\label{Eq_Ci}
 \end{align}
Plugging \eqref{vec_ti} into \eqref{er_i}, the RIS reflecting matrix that minimizes the total decoding error of the system is obtained by solving the following optimization problem
\begin{subequations}
\label{opt_initial}
\begin{align}
	 &\argmin_{\mathbf{w}_c} \sum_{i\in \mathcal{S}_s}\mathcal{F}(\mathbf{w}_c^H\mathbf{G}_i\mathbf{w}_c), \label{OptProb}\\
  &\mathrm{s.t.} \ \  \left|[\mathbf{w}_c]_{n}\right|=1, \label{constraint_unitModulus}\\
 & \ \ \ \ \ \ \mathbf{G}_i=\mathbf{C}_i(\mathbf{w}_c), \ i\in \mathcal{S}_s,
\end{align}
\end{subequations}
where $\mathbf{G}_i$ is an auxiliary parameter matrix. It is evident that neither the objective function in \eqref{OptProb} nor the constraint in \eqref{constraint_unitModulus} are convex. Therefore, it cannot be solved using the standard convex optimization solvers. However, in the following, we proceed to adapt and refine it prior to solving by two distinct algorithms, namely ASDR and AEVD. Details of these algorithms are delineated below.

\textbf{ASDR: } It can be proved that once $x$ surpasses a specific threshold, namely $\bar{\alpha}$, there is no further significant reduction in the value of $\mathcal{F}(x)$, and $\mathcal{F}(x)$ is a non-increasing function of $x$. Given this motivation, the problem in \eqref{opt_initial} can be approximated as
\begin{subequations}
\label{opt_initial2}
\begin{align}
	 \argmax_{\mathbf{w}_c} &\ \mathbf{w}_c^H\left(\sum_{i\in \mathcal{S}_s}\mathbf{G}_i\right)\mathbf{w}_c,\label{eqOpt2}\\\nonumber
  \mathrm{s.t.} &\left|[\mathbf{w}_c]_{n}\right|=1,\\
&  \mathbf{w}_c^H\mathbf{G}_i\mathbf{w}_c\leq \bar{\alpha}, \ \  i\in \mathcal{S}_s,\label{constraint_saturate}\\
 &  \mathbf{G}_i=\mathbf{C}_i(\mathbf{w}_c), \ i\in \mathcal{S}_s.
\end{align}
\end{subequations}
We solve this problem using a two-step alternating optimization method: In the first step, we estimate $\mathbf{G}_i$ by $  \mathbf{G}_i=\mathbf{C}_i(\mathbf{w}_{n-1})$, where $\mathbf{w}_{n-1}$ is the estimated RIS vector in the $(n-1)$th iteration, and in the second step, we solve the problem \eqref{opt_initial2} given $\mathbf{G}_i$. The details of the second step of the algorithm are as follows. We define $\mathbf{\bar{W}}=\mathbf{w}_c\mathbf{w}_c^H$, where $\mathbf{\bar{W}} \succeq 0$ and $\mathrm{rank}(\mathbf{\bar{W}})=1$, so we can write $\mathbf{w}_c^H\mathbf{G}_i\mathbf{w}_c=\mathrm{trace}(\mathbf{\bar{W}}\mathbf{G}_i)$. Since the rank-one constraint is non-convex, we relax it using SDR to obtain the following convex semidefinite program (SDP) from \eqref{opt_initial2}
\begin{align}
	 \argmax_{\mathbf{\bar{W}}} &\ \ \mathrm{trace}(\mathbf{\bar{W}}\mathbf{\bar{G}}),\label{eqOpt3}\\\nonumber
  \mathrm{s.t.} & \ [\mathbf{\bar{W}}]_{(n,n)}=1, \ n=1,...,Nn_s\\\nonumber
&  \mathrm{trace}(\mathbf{\bar{W}}\mathbf{G}_i)\leq \bar{\alpha},  i\in \mathcal{S}_s,
\end{align}
where $\mathbf{\bar{G}}=\sum_{i\in \mathcal{S}_s}\mathbf{G}_i$. The problem in \eqref{eqOpt3} can be solved using convex optimization solvers such as CVX. Note that the solution to the relaxed problem \eqref{eqOpt3} is not necessarily rank-1. Thus, we perform additional steps similar to \cite{Wu2018Intelligent} to ensure that the rank-1 constraint is satisfied. Particularly, by applying the eigenvalue decomposition (EVD) on the solution, $\mathbf{\bar{W}} = \mathbf{U}\mathbf{\Sigma}\mathbf{U}^H$, a sub-optimal solution to \eqref{eqOpt3} is obtained as
\begin{align}
\hat{\mathbf{w}}_c = \argmin_{l=1,...,T_{SDR}} \ \ \sum_{i\in \mathcal{S}_s}\mathcal{F}(\tilde{\mathbf{w}}_l^H\mathbf{G}_i\tilde{\mathbf{w}}_l),\label{rank1_application}
\end{align}
where $[\tilde{\mathbf{w}}_l]_{i}=[\bar{\mathbf{u}}_l]_{i}/\left|[\bar{\mathbf{u}}_l]_{i}\right|$, $\bar{\mathbf{u}}_l= \mathbf{U}\mathbf{\Sigma}^{1/2}\mathbf{r}_l$, $\mathbf{r}_l \sim \mathcal{CN}(\mathbf{0}, \mathbf{I}_{N^\prime})$ with $N^\prime =N$ and $N^\prime =Nn_s$ are selected for cases of $\mathcal{C}_0$ and $\mathcal{C}_1$, respectively, and $T_{SDR}$ denotes the number of realizations of $\mathbf{r}_l$. Note that for finding the solution in \eqref{rank1_application}, we minimize the main cost function in \eqref{opt_initial} instead of the approximated one in \eqref{opt_initial2}. The details of the ASDR method are shown in Algorithm \ref{algorithm1}, where $T_{\mathrm{iter}}$ is the total number of iterations.
   \begin{algorithm}
  \caption{ASDR method for RIS design.} 
  \label{algorithm1}
\textbf{Initialization}: $\mathbf{w}_0 = [e^{j\theta_1},...,e^{j\theta_{N^\prime }}]^T$, where $\theta_j\sim U(0,2\pi)$\\
			\For( {}){$n=1,2,...,T_{\mathrm{iter}}$}
   {
   \begin{enumerate}
   \item Calculate  $\mathbf{G}_i=\mathbf{C}_i(\mathbf{w}_{n-1})$ according to \eqref{Eq_Ci}.
   \item Calculate $\mathbf{\bar{W}}$ by SDP
   \item Perform EVD, $\mathbf{\bar{W}} = \mathbf{U}\mathbf{\Sigma}\mathbf{U}^H$.       
   \item Calculate $\mathbf{w}_n$ according to \eqref{rank1_application}.       
   \end{enumerate}
   }
 \end{algorithm}

   \begin{algorithm}
  \caption{AEVD method for RIS design.} 
  \label{algorithm2}
\textbf{Initialization}: $\mathbf{w}_0 = [e^{j\theta_1},...,e^{j\theta_{N^\prime}}]^T$, where $\theta_j\sim U(0,2\pi)$\\
			\For( {}){$n=1,2,...,T_{\mathrm{iter}}$}
			{
   1) Calculate  $\mathbf{G}_i=\mathbf{C}_i(\mathbf{w}_{n-1})$ according to \eqref{Eq_Ci}.\\
      2) Calculate $\mathbf{q}_{i,\max}$.\\
3) $\mathbf{v}_{n} =\dfrac{1}{|\mathcal{S}_s|}\sum_{i\in\mathcal{S}_s} \mathbf{q}_{i,\max}$.\\
    4) $[\mathbf{w}_{n}]_{i}=[\mathbf{v}_{n}]_{i}/\left|[\mathbf{v}_{n}]_{i}\right|,  i=1,...,N^\prime$\\
   5) $P_e(n)=\sum_{i\in \mathcal{D}_s}\mathcal{F}(\mathbf{w}_{n}^H\mathbf{G}_i\mathbf{w}_{n})$.\\
\If{$P_e(n)<\alpha_2$}{Stop the algorithm.}
   }
   \textbf{Output: } $\hat{\mathbf{w}}_c = \mathbf{w}_{n^\prime}$, where $n^\prime=\underset{n}{\text{argmin}}\  P_e(n)$
 \end{algorithm}

\textbf{AEVD: }To decrease the computational complexity of the RIS design algorithm, we propose AEVD for solving \eqref{opt_initial}. The AEVD method comprises the following steps: 1) First, we identify the RIS coefficient vectors that minimize the error for each user individually. To achieve this, we compute the eigenvector of the matrix $\mathbf{G}_i$ corresponding to its largest eigenvalue, denoted as $\mathbf{q}_{i,\max}$\footnote{For a Hermitian matrix $\mathbf{C}$ and a vector with unit norm, $\|\mathbf{w}\|^2=1$, the maximum value of $c=\mathbf{w}^H\mathbf{C}\mathbf{w}$ is obtained by choosing $\mathbf{w}$ as the eigenvector of $\mathbf{C}$ corresponding to its largest eigenvalue \cite{Lax2007Linear}.}, 2) subsequently, we calculate the average of these individual vectors to obtain a combined vector, 3) to ensure compliance with the unit-modulus constraint of the RIS elements, we rescale each element of the resulting vector. We repeat this AEVD algorithm for a total of $T_{\mathrm{iter}}$ iterations, stopping when $P_e(n)$ falls below the predefined threshold $\alpha_2$. A detailed description of the AEVD algorithm can be found in Algorithm \ref{algorithm2}.

\subsection{An Approximate Achievability Result}
\label{SecAchResult}
In this section, we establish an approximate achievability result for the RIS-aided URA scheme, considering the received signal model in \eqref{cascadedChannel}. For this purpose, we examine the coding scheme outlined below, while considering random coding. For encoding, every user maps its length-$B$ bit sequence to the rows of a $2^B\times n$ codebook, every element of which is drawn from $\mathcal{CN}(0,P^\prime)$. The PUPE of the system is defined as 
     \begin{align}
     \epsilon = p_{coll}+p_{cons} +  p_{mis}  \label{PUPE_theory}   ,
 \end{align}
where $p_{coll}$ denotes collision-related error, $p_{\text{cons}}$ is the error stemming from the average transmitted power surpassing the power constraint, and $p_{mis}$ is the ratio of the expected number of undetected users to the total number of active users. The first term is obtained by considering the events that 2 out of $K_a$ users select the same codeword, which satisfies
\begin{align}
    p_{coll} \leq  \dfrac{\binom{K_a}{2}}{2^{B}},
\end{align}
The second term is due to the event that the norm squared of at least one user's codeword, which follows a chi-squared distribution, is greater than $nP$:
\begin{align}
    p_{cons} \leq K_a\left(1-F_{2n}\left({{2nP}/{P^\prime}}\right)\right),
\end{align}
where $F_{2n}(x)$ is the cumulative distribution function of the chi-squared distribution with $2n$ degrees of freedom. To determine $p_{mis}$, we need to conduct a more detailed analysis as given below. Note that the assumptions put forth in the following analysis are consistent with the achievable rate calculation, since they result in a rise in the decoding error.

\textbf{Calculation of an approximate upper bound for $\boldsymbol{p_{miss}}$}:
 Considering Gaussian codewords, the received signal in \eqref{cascadedChannel} can be written as
\begin{align}
       \mathbf{Y} &=  \sum_{i=1}^{L_T}\mu_i\mathbf{G}\mathbf{O}_i +\mathbf{Z}, \label{eq57}      
\end{align}
where $ \mathbf{Y}\in \mathbb{C}^{M\times n}$, $\mathbf{O}_i=\sqrt{N}\mathrm{diag}(\bar{\mathbf{a}}_{N,\bar{\phi}_i,\bar{\psi}_i})\mathbf{W}\mathrm{diag}(\mathbf{p}_i)$, $L_T=\sum_{l=1}^{K_a}L_{R,i}$, $\mathbf{p}_i$ is the Gaussian codeword transmitted through the $i$th path (the same codeword may be transmitted via multiple paths), and $\mathbf{W}\in\mathbb{C}^{N\times n} $ is the RIS phase shift matrix with unit modulus elements whose phases are uniformly distributed between $0$ and $2\pi$. Since randomly shifting the phase of a complex and circularly symmetric Gaussian random variable does not change its probability distribution function, we have $[\mathbf{O}_i]_{k,l}\sim \mathcal{CN}\left(0,P^\prime\right)$. Vectorizing both sides of \eqref{eq57}, we get
\begin{align}
    \mathbf{y} &= \sum_{i=1}^{L_T}\mu_i\mathbf{q}_i+ \mathbf{z}, \label{received_gasuuy}
\end{align} 
where $ \mathbf{y}\in \mathbb{C}^{Mn\times 1}$, $\mathbf{q}_i=\mathbf{G}^\prime\mathbf{o}_i\sim \mathcal{CN}\left(\mathbf{0},\Sigma\right)$ with $\Sigma=P^\prime\mathbf{G}^\prime{\mathbf{G}^\prime}^H$, $\mathbf{o}_i=\mathrm{vec}\left(\mathbf{O}_i\right)$, and $\mathbf{G}^\prime=(\mathbf{I}_n\otimes \mathbf{G})$. To detect active scatterers, namely $\mathbf{q}_i$'s for $i=1,2,...,L_T$, an iterative algorithm is employed. At each iteration, the scatterer with the highest energy is detected, its corresponding path gain $\mu_i$ is estimated, and its contribution is removed from the received signal. Thus, in the $k$th iteration of the algorithm, $k-1$ scatterers have been detected and removed. We assume that as soon as a false scatterer is detected in an iteration, the algorithm halts. Therefore, the analysis is pessimistic and gives the achievability result. The $k$th iteration of the algorithm is detailed below:

The received signal vector in \eqref{received_gasuuy} can be restated in the following form
\begin{align}
    \mathbf{y} &= \mathbf{Q}_k\mathbf{M}_k+\sum_{j=\mathcal{L}_k}\mu_j\mathbf{q}_j+ \mathbf{z},
\end{align} 
where columns of $\mathbf{Q}_k \in \mathbb{C}^{nM\times (k-1)}$ and rows of $\mathbf{M}_k \in \mathbb{C}^{(k-1)\times 1}$ are $\mathbf{q}_i$'s and $\mu_i$'s corresponding to $k-1$ detected scatterers, and $\mathcal{L}_k$ is the set of undetected scatterers' indices, with $|\mathcal{L}_k|=L_T-k+1$. By employing LS, $\mathbf{M}_k$ can be estimated as
\begin{align}
    \hat{\mathbf{M}}_k=(\mathbf{Q}_k^H\mathbf{Q}_k)^{-1}\mathbf{Q}_k^H\mathbf{y} .
\end{align}
Having $\hat{\mathbf{M}}_k$ and $\mathbf{Q}_k$, the contribution of detected scatterers can be removed from the received signal by subtraction. Thus, the remaining received signal in the $k$th iteration can be written as
\begin{align}
\nonumber
    \mathbf{y}_k &=\mathbf{y}-\mathbf{Q}_k\hat{\mathbf{M}}_k\\
    &=\sum_{j=\mathcal{L}_k}\mu_j\bar{\mathbf{Q}}\mathbf{q}_j+ \bar{\mathbf{Q}}\mathbf{z},\label{eq44_aug12}
\end{align} 
where $\bar{\mathbf{Q}}=\mathbf{I}_{Mn_s}-\mathbf{Q}_k(\mathbf{Q}_k^H\mathbf{Q}_k)^{-1}\mathbf{Q}_k^H$. Note that for obtaining \eqref{eq44_aug12}, we use the property $\bar{\mathbf{Q}}\mathbf{Q}_k=\mathbf{0}$. Since the entries of \( \mathbf{Q}_k \), \( \mathbf{q}_j \), and \( \mathbf{z} \) are independent and have zero mean, and assuming that \( nM \) is large, by the Weak Law of Large Numbers, we have \(\mathbf{Q}_k^H \mathbf{q}_j \approx\mathbf{Q}_k^H \mathbf{z} \approx \mathbf{0} \). Therefore, an approximation of $\mathbf{y}_k$ in \eqref{eq44_aug12} can be attained as
\begin{align}
      \nonumber
      \mathbf{y}_{k} &\approx \sum_{j=\mathcal{L}_k}\mu_j\mathbf{q}_j+ \mathbf{z}\\
        &= \mu_i\mathbf{q}_i+\bar{\mathbf{y}}_{k,i} ,  i\in\mathcal{L}_k,\label{Signal_intNoise}
\end{align} 
where the first and second terms in the right-hand side are the desired signal term for the $i$th user and the interference-plus-noise term, respectively, and $\bar{\mathbf{y}}_{k,i} =  \sum_{j\in \mathcal{L}_k\backslash i}\mu_j\mathbf{q}_j+ \mathbf{z}$. 
 Let $\mathcal{C}_T$ be the set of all possible scatterers from the users to the RIS. In the $k$th iteration of the algorithm, the receiver detects the scatterer with highest energy by solving
\begin{align}
    i=\max_{u\in \mathcal{U},\bar{i}\in \mathcal{C}_T}\ \zeta_{k\bar{i}u},
\end{align}
where $\zeta_{kiu}=\mathrm{Re}\left(u  \mathbf{y}_{k}^H\mathbf{q}_{i}\right)$, $\mathcal{U}=\left\{\pm {1}/{\sqrt{2}} \pm {1}/{\sqrt{2}}\right\}$. Given that the algorithm halts when a false scatterer ($\mathbf{q}_j$ for $j \notin \mathcal{L}_k$) is detected, the probability of the algorithm stopping in the $k$th iteration is expressed as
\begin{subequations}
\begin{align}
    P_{stop,k} &= \mathbb{P}\left(\bigcup_{\tiny \begin{matrix}q\notin\mathcal{L}_k\\u_2\in \mathcal{U}\end{matrix}}\bigcap_{\tiny \begin{matrix}j\in \mathcal{L}_k\\u_1\in \mathcal{U}\end{matrix}}\{\zeta_{kju_1}<\zeta_{kqu_2}\}\right) \\
&\leq  {\scriptsize \left(\sum_{\tiny \begin{matrix}q\notin\mathcal{L}_k\\u_2\in \mathcal{U}\end{matrix}}\ \mathbb{P}\left(\bigcap_{\tiny \begin{matrix}j\in \mathcal{L}_k\\u_1\in \mathcal{U}\end{matrix}}\{\zeta_{kju_1}<\zeta_{kqu_2}\}\right) \right)^\rho}\label{eqq68}\\
&\leq  {\scriptsize \left(\sum_{\tiny \begin{matrix}q\notin\mathcal{L}_k\\u_2\in \mathcal{U}\end{matrix}}\ \mathbb{P}\left(\zeta_{ki\bar{u}}<\zeta_{kqu_2}\right) \right)^\rho}, i=\max( \mathcal{L}_k)\label{eqq698}\\
&=  {\scriptsize \left(\sum_{\tiny \begin{matrix}q\notin\mathcal{L}_k\\u_2\in \mathcal{U}\end{matrix}}\ p_{\epsilon} \right)^\rho}\label{47d}, 
\end{align}
\label{FalseProb}
\end{subequations}
\noindent where 
\begin{align}
    p_{\epsilon}=\mathbb{P}\left(\zeta_{ki\bar{u}}<\zeta_{kqu_2}\right),\label{70_21_04}
\end{align}
with $i=\max(\mathcal{L}_k)$, $ q\notin\mathcal{L}_k,u_2\in \mathcal{U}$, $\bar{u}=\argmax_{u\in \mathcal{U}} (\mathrm{Re}(u\mu_{k}^*))$. In \eqref{eqq68} we use Gallager’s $\rho$-trick which states that $\mathbb{P}\left(\bigcup_{i\in \mathcal{S}} A_i\right) \leq (\sum_{i\in \mathcal{S}} \mathbb{P}\left(A_i\right))^\rho$ for $0\leq\rho\leq 1$. Additionally, in \eqref{eqq698}, we use the fact $\mathbb{P}\left(\bigcap_{i\in \mathcal{S}} A_i\right) \leq  \mathbb{P}\left(A_j\right)$ for $\  j \in\mathcal{S}$. Using the received signal model in \eqref{Signal_intNoise} and the property $\mathbb{P}\left(x>0\right)\leq\mathbb{E}\{e^{\lambda x}\}$ for $\lambda>0$, \eqref{70_21_04} can be written as
\begin{align}
     p_{\epsilon}&\leq  \mathbb{E}_{\mathcal{G}}\left\{e^{\lambda E_{y,i,q}}\right\} \label{eq72},
\end{align}
where $ E_{y,i,q}=-\zeta_{ki\bar{u}}+\zeta_{kqu_2}=-\mathrm{Re}\left(\bar{u}\mu_i^*\right)\mathbf{q}_i^H\mathbf{q}_i-\mathrm{Re}\left(\bar{u}\bar{\mathbf{y}}_{k,i}^H\mathbf{q}_i\right)+\mathrm{Re}\left(u_2\mu_i^*\mathbf{q}_i^H\mathbf{q}_q\right)+\mathrm{Re}\left(u_2\bar{\mathbf{y}}_{k,i}^H\mathbf{q}_q\right)$, and $\mathcal{G}=\{\bar{\mathbf{y}}_{k,i},\mathbf{q}_i,\mathbf{q}_q\}$.  
\begin{mydef3}
\label{Lemm1}
For $\mathbf{a}\sim \mathcal{CN}\left(\mathbf{0},\mathbf{A}\right)$, $\mathbf{r}\in \mathbb{C}^{1\times K}$, and $\mathbf{A},\mathbf{B}\in \mathbb{C}^{K\times K}$, if $\mathbf{I}_K-\mathbf{A}\mathbf{B}\succcurlyeq 0$, we have 
\begin{align}
    \mathbb{E}_{\mathbf{a}}\left\{    e^{ \mathbf{a}^H \mathbf{B}\mathbf{a}+ \mathrm{Re}\left(\mathbf{r}\mathbf{a}\right)}\right\}= \dfrac{1}{ |\mathbf{C}|}     e^{0.25  \mathbf{r}\mathbf{A}\mathbf{C}^{-1}\mathbf{r}^H},
\end{align}
where $\mathbf{C}= (\mathbf{I}_K-\mathbf{A}\mathbf{B})$.
\end{mydef3}
\begin{proof}
Let $v=  \mathbb{E}_{\mathbf{a}}\left\{    e^{ \mathbf{a}^H \mathbf{B}\mathbf{a}+ \mathrm{Re}\left(\mathbf{r}\mathbf{a}\right)}\right\}$. We have
\begin{align}
\nonumber v  &  =  \int \dfrac{1}{ |\pi\mathbf{A}|}e^{-\mathbf{a}^H\mathbf{A}^{-1}\mathbf{a}}\ \     e^{ \mathbf{a}^H \mathbf{B}\mathbf{a}+ \mathrm{Re}\left(\mathbf{r}\mathbf{a}\right)} d\mathbf{a}\\\nonumber
      &= \dfrac{|\pi\mathbf{D}|}{ |\pi\mathbf{A}|} \int \dfrac{1}{ |\pi\mathbf{D}|}e^{-\mathbf{a}^H\mathbf{D}^{-1}\mathbf{a}}\ \     e^{+ \mathrm{Re}\left(\mathbf{r}\mathbf{a}\right)} d\mathbf{a}\\  
           &= { |\mathbf{I}_K-\mathbf{A}\mathbf{B}|}^{-1}     e^{0.25  \mathbf{r}\mathbf{D}\mathbf{r}^H}  \label{eqs56}
\end{align}
where $\mathbf{D}=  \mathbf{A}(\mathbf{I}_K-\mathbf{A}\mathbf{B})^{-1}$. In \eqref{eqs56}, we leverage the characteristic property of the moment generation function associated with the Gaussian distribution.
\end{proof}
\begin{figure}[t!]%[h!]
	\centering
	\includegraphics[width=1\linewidth]{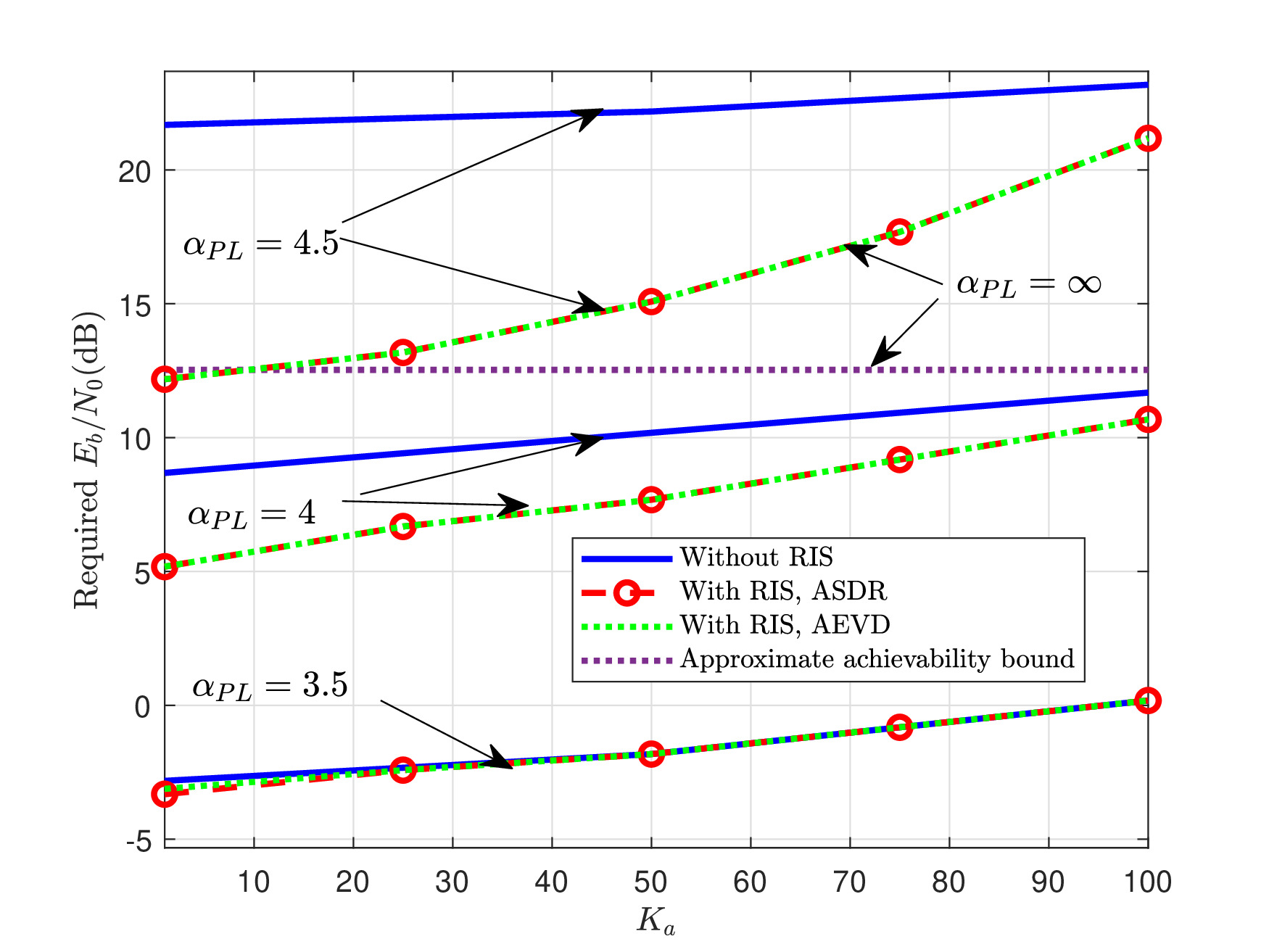}
	\caption{{\small The required $E_b/N_0$ to achieve the target PUPE of $0.1$ for different user-BS path-loss exponents.}	}
	\label{FIG_WithoutRIS}
\end{figure}
 Since $\mathbf{q}_q$, $ \mathbf{q}_i$, and $\bar{\mathbf{y}}_{k,k}$ are independent of each other, we can use the law of total expectation to compute the expectation in \eqref{eq72}. Initially, we compute the expectation with respect to $\bar{\mathbf{y}}_{k,k}$ while keeping other variables fixed. Using Lemma \ref{Lemm1} and considering that $\bar{\mathbf{y}}_{k,k}\sim \mathcal{CN}\left(\mathbf{0},\Sigma_y\right)$, where $\Sigma_y=\delta^2_{k}\Sigma+\sigma_z^2\mathbf{I}_{Mn}$, we get 
 \begin{align}
     \mathbb{E}_{\bar{\mathbf{y}}_{k,k}}\left\{e^{\lambda E_{y,i,q}}\right\}=  e^{E^{\prime}_{i,q}},\label{82_25_4}
 \end{align}
where $\delta^2_{k}=\sum_{j\in \mathcal{L}_k}|\mu_j|^2$, $E^{\prime}_{i,q}=-\mathrm{Re}\left(\bar{u}\mu_i^*\right)\lambda \mathbf{q}_i^H\mathbf{q}_i+\mathrm{Re}\left(0.25 \lambda^2   \mathbf{q}_i^H\Sigma_y  \mathbf{q}_i\right)+  \mathbf{q}_q^H0.25 \lambda^2  \Sigma_y  \mathbf{q}_q+\mathrm{Re}\left(\mathbf{q}_i^H \mathbf{A}_1\mathbf{q}_q\right)$, $\mathbf{A}_1 =  (u_2 \bar{u}^*s_{i1}-2s_2 \bar{u}^*u_2 \sigma_z^2) \mathbf{I}_{Mn}+ s_2 \bar{u}^*u_2 \delta^2_{k} \Sigma $, $s_{i1} = \lambda \mathrm{Re}(\bar{u}\mu_i^*)$, and $ s_2=-0.5\lambda^2$.\\
\indent Then, by taking expectation with respect to $\mathbf{q}_q$ and using Lemma \ref{Lemm1} we have 
  \begin{align}
      \mathbb{E}_{\mathbf{q}_q}\{e^{E^{\prime}_{i,q}}\}=\dfrac{1}{ |\mathbf{C}_2|}e^{\mathrm{Re}\left(\mathbf{q}_i^H\mathbf{C}_1\mathbf{q}_i\right)},\label{74_21_04}
  \end{align}
    where $\mathbf{C}_1=-\lambda \mathrm{Re}(\bar{u}\mu_i^*)+  0.25 \lambda^2\Sigma_y  + 0.25\mathbf{A}_1\Sigma\mathbf{C}_2 ^{-1} \mathbf{A}_1^H$, and $ \mathbf{C}_2=\mathbf{I}_{Mn} -0.25 \lambda^2\sigma_z^2 \Sigma-0.25 \lambda^2\delta^2_{k}\Sigma^2$. To ensure $\mathbf{C}_2\succcurlyeq 0$ as the necessary condition of Lemma \ref{Lemm1}, the parameter $\lambda$ needs to be chosen such that all eigenvalues of $\mathbf{C}_2$ are positive. Leveraging the properties of eigenvalues\footnote{If $b$ is an eigenvalue of the matrix $\mathbf{B}$, then the corresponding eigenvalue of $\sum_{i=1}^N a_i\mathbf{B}^i$ is $\sum_{i=1}^N a_ib^i$, and the eigenvalues of $\mathbf{I}_{n}\otimes \mathbf{B}$ are simply the eigenvalues of $\mathbf{B}$ repeated $n$ times.
}, the eigenvalues of $\mathbf{C}_2$ are calculated as $1 -0.25\lambda^2\left( \sigma_z^2 \sigma_{G_j}+ \delta^2_{k}\sigma_{G_j}^2\right)$, where $\sigma_{G_j}$ is the $j$th eigenvalue of the matrix $ P^\prime\mathbf{G}\mathbf{G}^H$. Thus, the parameter $\lambda$ is valid in the range $0\leq\lambda<\lambda_0$, where $\lambda_0=2/\sqrt{  \sigma_{\max}\sigma_z^2+\delta^2_{k}\sigma_{\max}^2}$ and $\sigma_{\max}$ is the maximum of $\sigma_{G_j}$'s.

Finally, taking expectation from the result in \eqref{74_21_04}, with respect to $\mathbf{q}_i$, and using Lemma \ref{Lemm1}, we get
\begin{subequations}
    \begin{align}
\mathbb{E}_{\mathcal{G}}\left\{e^{\lambda E_{y,i,q}}\right\} &\leq \dfrac{1}{ |\mathbf{C}_2| |\mathbf{I}_n-\Sigma\mathbf{C}_1|} \\
&=\dfrac{1}{ |\mathbf{C}_2-\mathbf{C}_2\Sigma\mathbf{C}_1|} \label{eq71}\\
&=\dfrac{1}{|\mathbf{I}_{Mn} +c_{i1} \Sigma+c_{ik2} \Sigma^2|}\label{eqqq72s}\\
&=\dfrac{1}{\left(\prod_{j=1}^{M}\left(1 +c_{i1} \sigma_{G_j}+c_{ik2} \sigma_{G_j}^2\right)\right)^n}\label{85d}\\
&=\epsilon_{\lambda,k,i}\label{85e},
 \end{align}
\label{eqqq72}
\end{subequations}
\noindent where $\epsilon_{\lambda,k,i}=e^{-n\log\left(\prod_{j=1}^{M}\left(1 +c_{i1} \sigma_{G_j}+c_{ik2} \sigma_{G_j}^2\right)\right)}$, $c_{i1} = s_2 \sigma_z^2+s_{i1}$ and $c_{ik2} =  s_2 \delta^2_{k}-0.25s_{i1}^2$. In \eqref{eq71}, we use the determinant product rule; In \eqref{eqqq72s}, we perform some calculations, considering that $\mathbf{A}_1$, $\mathbf{C}_2$, and $\Sigma$ are commutative with respect to multiplication; \eqref{85d} is derived from the properties of the eigenvalue.

In the $k$th iteration, there are $\binom{L_T}{k-1}$ possible sets for $\mathcal{L}_k$. To meet the achievability assumption, we choose the set that gives the highest error. Concentrating on \eqref{eqqq72}, we can interpret that the upper bound for $\mathbb{E}_{\mathcal{G}}\left\{e^{\lambda E_{y,i,q}}\right\}$ is an increasing function of $\delta_k^2$. We assume without loss of generality that $|\mu_1|\leq...\leq|\mu_{L_T}|$. Looking at the structure of the $\delta_k^2$ below \eqref{82_25_4}, and since $i= \max(\mathcal{L}_k)$, choosing $\mathcal{L}_k = i-|\mathcal{L}_k|+1,...,i$ gives the maximum $\delta_y^2$, hence increasing the upper bound of $\mathbb{E}_{\mathcal{G}}\left\{e^{\lambda E_{y,i,q}}\right\}$. The desired scatterer index $i$ that gives largest error is selected by the following search 
\begin{align}
    i=\argmax_{i^\prime:\{|\mathcal{L}_k|,|\mathcal{L}_k|+1, ..., L_T\}} \epsilon_{\lambda,k,i^\prime}.\label{85_25_04}
\end{align}
Focusing on  \eqref{47d}, \eqref{eq72}, and \eqref{85e}, and considering that $|\mathcal{C}_T|=2^BN$, $|\mathcal{U}|=4$, $|\mathcal{L}_k|=L_T-k+1$, an approximate upper bound for $P_{stop,k}$ is obtained as
\begin{align}
    P_{stop,k}&\leq    \argmin_{\tiny \begin{matrix}
    0\leq\lambda<1/\lambda_0\\\rho\in\{0,1\}
\end{matrix}}  e^{ \rho\left((2+B)\log2+\log N+ \epsilon_{\lambda,k,i}\right)}.\label{Eq57Aug12}
\end{align}
If the algorithm stops in the $k$th iteration (it detects true scatterers in the initial $k-1$ iterations and then detects a false scatterer in the $k$th iteration), $p_{miss}$ becomes $1-K_k/K_a$, where $K_k$ is the number of users to which the detected scatterers belong. Approximately assuming that the algorithm performs independently from the previous iterations, the probability that it stops at the $k$th iteration is approximated as $P_{stop,k}\prod_{l=1}^{k-1}(1- P_{stop,l})$. Then the expected $p_{miss}$ is calculated as
\begin{align}
   p_{mis}\approx (1-K_k/K_a) P_{stop,k}\prod_{l=1}^{k-1}(1- P_{stop,l}).\label{PStopss}
\end{align}
Note that \(p_{miss}\) is derived for a given set of path gains, $\mathcal{M}=\{\mu_1,\mu_2,...,\mu_{L_T}\}$. Therefore, by applying expectation across $\mathcal{M}$, the ultimate bound is determined, which remains unaffected by the path gains. Given the analytical challenge in taking expectation with respect to $\mathcal{M}$, we generate multiple realizations to perform their evaluation via Monte Carlo method.

\section{RISUMA with Direct User-BS Link}
\label{Sec54}
In this section, we extend the proposed RISUMA approach by modifying it to accommodate scenarios where also a direct path exists between the users and the BS. In this case, the system model is the same as the one in Section \ref{SystemModel}, except for the existence of the direct paths between the users and the BS. Hence, the received signal at the BS can be modeled as
\begin{align}
\nonumber
    \mathbf{y}_t = \sum_{i=1}^{K_a} \left(\mathbf{G}\mathrm{diag}(\mathbf{h}_i)\mathbf{w}_t + \mathbf{d}_i \right)x_{i,t}&+\mathbf{z}_t, \in \mathbb{C}^{M\times 1} \\
    &, \ t=1,...,n, \label{received_complete}
\end{align}
where $\mathbf{d}_i$ is modeled as
\begin{align}
\mathbf{d}_i = \sqrt{M} \sum_{g_i =1}^{ L_{B,i} }\mu_{g_i}{\mathbf{a}_{M}(\phi_{i,g_i},\psi_{i,g_i})}^T \in \mathbb{C}^{M\times 1},\label{eqs4}
\end{align}
where $L_{B,i}$ is the number of direct paths between the $i$th user and the BS, and $(\phi_{i,g_i},\psi_{i,g_i})$ are azimuth and elevation AOAs at the BS for the $g_i$th direct path of the $i$th user's signal. For the user-BS direct path, we consider the same path-loss model as the RIS-BS and the user-RIS channels in Section \ref{SystemModel}, with $d_{g_i}$ and $\mu_{g_i}$ being the distance and path gain of the $g_i$th direct path, respectively. Employing the same encoder as in Section \ref{encoder}, while considering the user-BS direct channel, the received signal models in \eqref{receivedP} and \eqref{receivedc} are modified as

% $\mathbf{Y}_{p} = \sqrt{ P_p}\sum_{i\in\mathcal{S}_s}^{}\mathbf{g}_i\mathbf{W}_{p_s}\mathrm{diag}(\mathbf{p}_i)+\mathbf{d}_i\mathbf{p}_i+ \mathbf{Z}_{p}
% $, and $\mathbf{Y}_{c,f} = \sqrt{ P_c}\sum_{i\in\mathcal{S}_s}^{}  \left(\mathbf{g}_i\mathbf{W}_{c_s}\mathrm{diag}(\mathbf{b}_i) +\mathbf{d}_i\mathbf{b}_i\right)v_{i,f}+\mathbf{Z}_{c,f}$, where $\mathbf{g}_i=\mathbf{G}\mathrm{diag}(\mathbf{h}_{i} )$.
\begin{subequations}
\label{received_Yc_Multi}
\begin{align}
\mathbf{Y}_{p} &= \sqrt{ P_p}\sum_{i\in\mathcal{S}_s}^{}\mathbf{G}\mathrm{diag}(\mathbf{h}_{i} )\mathbf{W}_{p_s}\mathrm{diag}(\mathbf{p}_i)+\mathbf{d}_i\mathbf{p}_i+ \mathbf{Z}_{p},
 \end{align}
 \begin{align}
\nonumber
\mathbf{Y}_{c,f} &= \sqrt{ P_c}\sum_{i\in\mathcal{S}_s}^{}  \left(\mathbf{G}\mathrm{diag}(\mathbf{h}_{i} )\mathbf{W}_{c_s}\mathrm{diag}(\mathbf{b}_i) +\mathbf{d}_i\mathbf{b}_i\right)v_{i,f}\\
   & \indent \indent \indent \indent \indent \indent \indent \indent \indent \indent \indent \indent \indent \indent \indent \indent +\mathbf{Z}_{c,f}.  
 \end{align}
\end{subequations}
\subsection{Decoder Structure}
The decoder consists of a joint pilot detector and channel estimator algorithm, data detection part, and RIS design, whose details are described below.
\subsubsection{Joint Pilot Detection and Channel Estimation}
The received signal matrices in \eqref{received_Yc_Multi} resemble the ones for the scenario of blocked user-BS channels in \eqref{receivedP} and \eqref{receivedc}, with the addition of an extra term stemming from the direct user-BS channels, denoted as $\mathbf{d}_i$. Thus, we modify the JDCE algorithm in Section \ref{Sec_ch_estimation} to estimate $\mathbf{d}_i$ as well as the user-RIS channel $\mathbf{h}_i$. Different steps of the modified version are described below.
\begin{itemize}
\item \textbf{Step 1} (Pilot detection): the pilot with highest energy (with index $\hat{k}$) is obtained as
\begin{align}
\hat{k}&= \max(\hat{k}_1,\hat{k}_2),\label{energyDetector_direct}\\
    \hat{k}_1&=\argmax_{k\in{1,2,...,2^{B_p}}} \dfrac{\mathrm{trace}(\mathbf{Q}_k\mathbf{R}_{pj} \mathbf{Q}_k^H)}{\mathrm{trace}(\mathbf{Q}_k\mathbf{Q}_k^H)},\\
  \hat{k}_2&=\argmax_{k\in{1,2,...,2^{B_p}}} \dfrac{\bar{\mathbf{p}}_k\mathbf{R}_{pj} \bar{\mathbf{p}}_k^H}{\|\bar{\mathbf{p}}_k\|^2}.    
\end{align}
Note that if $\hat{k}_1>\hat{k}_2$, the detected scatterer belongs to the user-RIS-BS channel; otherwise, it belongs to the user-BS direct channel.

\item \textbf{Step 2} (path detection): using the following formulation, we find the AOAs of the $\hat{k}$th pilot's strongest scatterer:
\begin{align}
    (\hat{l},\hat{q})=\argmax_{l\in \mathcal{T}(N_1),q\in \mathcal{T}(N_2)} \delta_{l,q}, \label{pathDetect_Direct}
\end{align}
\begin{align}
    \delta_{l,q}= \left\{\begin{matrix}
 \dfrac{\mathrm{trace}(\mathbf{F}_{l,q}\mathbf{R}_{pj} \mathbf{F}_{l,q}^H)}{\mathrm{trace}(\mathbf{F}_{l,q}\mathbf{F}_{l,q}^H)}   \ \ \  & \mathrm{if } \ \ \  \hat{k}=\hat{k}_1\\ 
 \bar{\mathbf{a}}_{N,l,q} ^*\mathbf{Y}_{p}^j\bar{\mathbf{p}}_{\hat{k}}^H   \ \ \  & \mathrm{if } \ \ \ \hat{k}=\hat{k}_2
\end{matrix}\right..
\end{align}
\item \textbf{Step 3 }(SIC): Using 
\begin{align}
    [\mathbf{\bar{U}}]_{(:,j)}=\left\{\begin{matrix} \sqrt{P_p}\mathbf{F}_{\hat{l},\hat{q}} \ \ \ \ \ \ &  \mathrm{if } \ \ \  \hat{k}=\hat{k}_1\\
     \sqrt{P_p}\bar{\mathbf{a}}_{N,\hat{l},\hat{q}} ^T\bar{\mathbf{p}}_{\hat{k}} &  \mathrm{if } \ \ \  \hat{k}=\hat{k}_1\end{matrix}\right.
\end{align}
 instead of the one in Section \ref{Sec_ch_estimation}, the SIC operation in the $j$th iteration can be performed as in \eqref{SIC_blocked}. The stopping criterion remains the same as the one in \eqref{StoppingDetector}.

\end{itemize}
\subsubsection{Data Detection}
\label{sec_polar_multi} We employ the same MMSE-based decoder as in Section \ref{sec_polar}, except for the modified matrix $\mathbf{T}_i$:
\begin{align}
\mathbf{T}_i  =  \sqrt{ P_c}\left(\mathbf{G}\mathrm{diag}(\mathbf{h}_i )\mathbf{W}_{c_s} \mathrm{diag}(\mathbf{b}_i)+\mathbf{d}_i\mathbf{b}_i\right)\in \mathbb{C}^{M\times n_s} , \label{T_i_direct}
\end{align}
to account for the direct path.
\subsubsection{RIS Design}
In a similar way as in Section \ref{SectionRISdesign}, the SINR of the $i$th user in the input of the polar decoder is calculated as $\beta_i = \sigma_{s,i}/(1-\sigma_{s,i})$ where $\sigma_{s,i}=    \mathbf{t}_i^H   \mathbf{\hat{R}}^{-1}\mathbf{t}_i$, and $\mathbf{t}_i=\mathrm{vec}\left(\mathbf{T}_i\right)$. Then, using \eqref{T_i_direct}, we can write $    \mathbf{t}_i  =  \mathbf{E}_i\mathbf{w}_c+\mathbf{e}_i $, where $\mathbf{E}_i$ and $\mathbf{w}_c$ are defined in \eqref{W_c_structures2} and \eqref{W_c_structures3}, respectively, and $\mathbf{e}_i=\mathrm{vec}(\mathbf{d}_i\mathbf{b}_i)$. Therefore, we can write $\sigma_{s,i}$ in the following form
\begin{align}
\nonumber
\sigma_{s,i}  =& \   \mathbf{w}_c^H\mathbf{E}_i^H \mathbf{\hat{R}}^{-1}\mathbf{E}_i\mathbf{w}_c+\mathbf{w}_c^H\mathbf{E}_i^H\mathbf{\hat{R}}^{-1}\mathbf{e}_i\\\nonumber&+\mathbf{e}_i^H\mathbf{\hat{R}}^{-1}\mathbf{E}_i\mathbf{w}_c+\mathbf{e}_i^H\mathbf{\hat{R}}^{-1}\mathbf{e}_i, \\
=& \  \bar{\mathbf{w}}_c^H\mathbf{C}_i^\prime(\bar{\mathbf{w}}_c)\bar{\mathbf{w}}_c \ \label{Sigma_direct}
\end{align}
 where
\begin{align}
     \mathbf{C}_i^\prime (\bar{\mathbf{w}}_c)&=\left[\begin{matrix}\mathbf{E}_i^H \mathbf{\hat{R}}^{-1}\mathbf{E}_i&\mathbf{E}_i^H\mathbf{\hat{R}}^{-1}\mathbf{e}_i\\\mathbf{e}_i^H\mathbf{\hat{R}}^{-1}\mathbf{E}_i&\mathbf{e}_i^H\mathbf{\hat{R}}^{-1}\mathbf{e}_i\end{matrix}\right], \label{C_prime}\\
     \bar{\mathbf{w}}_c&=\left[\begin{matrix}\mathbf{w}_c \\ 1
     \end{matrix}\right].
     \end{align}
We can see that the parameter $\sigma_{s,i}$ has exactly the same structures as in \eqref{Sigma_direct} and \eqref{vec_ti}. Hence, following the same argument as in Section \ref{SectionRISdesign}, the RIS reflecting matrix that minimizes the total decoding error of the system is obtained by solving the following optimization problem:
\begin{subequations}
\label{opt_initial3}
\begin{align}
	 &\argmin_{ \bar{\mathbf{w}}_c} \sum_{i\in \mathcal{S}_s}\mathcal{F}( \bar{\mathbf{w}}_c^H\mathbf{G}_i \bar{\mathbf{w}}_c), \label{OptProb2}\\
  &\mathrm{s.t.} \ \  \left|[ \bar{\mathbf{w}}_c]_{n}\right|=1, \ n=1,2,...,N^\prime \label{constraint_unitModulus2}\\
 & \  \ \ \ \ \ \ [ \bar{\mathbf{w}}_c]_{N^\prime+1}=1, \label{51c}\\
 & \ \ \ \ \ \ \mathbf{G}_i=\mathbf{C}_i^\prime( \bar{\mathbf{w}}_c), \ i\in \mathcal{S}_s,
\end{align}
\end{subequations}
where $\mathbf{G}_i$ is an auxiliary parameter matrix. The problem in \eqref{opt_initial3} can be solved via the modified ASDR and AEVD algorithms as described below.

\textbf{ASDR:} The modified ASDR algorithm is obtained by revising Algorithm \ref{algorithm1}. Particularly, in the first step of Algorithm \ref{algorithm1}, we calculate $\mathbf{G}_i=\mathbf{C}_i^\prime(\mathbf{w}_{n-1})$ according to \eqref{C_prime}. For the SDP in the second step, we replace $N^\prime+1$ with $N^\prime $. For the fourth step, $\hat{\mathbf{w}}_c$ is obtained by

\begin{align}
\hat{\mathbf{w}}_c = \argmin_{l=1,...,T_{SDR}} \sum_{i\in \mathcal{S}_s}\mathcal{F}\left(\begin{bmatrix}\tilde{\mathbf{w}}_l\\1\end{bmatrix}^H\mathbf{G}_i\begin{bmatrix}\tilde{\mathbf{w}}_l\\1\end{bmatrix}\right).
\end{align}
\textbf{AEVD:} The modified AEVD is the same as Algorithm \ref{algorithm2}, except replacing $N^\prime+1$ with $N^\prime$, and correcting the first and second steps as follows. In the first step, the $\mathbf{G}_i$ matrix is obtained by $\mathbf{G}_i = \mathbf{C}_i^\prime (\bar{\mathbf{w}}_{n-1})$. In the second step, we find the eigenvector corresponding to the largest eigenvector, $\mathbf{q}_{i,\max}\in \mathbb{C}^{(N^\prime+1)\times 1}$, then we calculate $\mathbf{q}_{i,\max}=\dfrac{\mathbf{q}_{i,\max}}{[\mathbf{q}_{i,\max}]_{N^\prime+1}}$ to satisfy the constraint in \eqref{51c}.

\subsection{Computational Complexity}
\label{Section_COmputaional}
In this part, we assess the computational complexity of the proposed RISUMA approach, by considering the number of multiplications as a measure of computational complexity.
\subsubsection{Joint Pilot Detection and Channel Estimation} Energy detectors in \eqref{energyDetector} and \eqref{energyDetector_direct} has a computational complexity of $\mathcal{O}\left(MNn_p2^{B_p}\right)$, with $\mathcal{O}(.)$ referring to the standard big-O notation, denoting the order of complexity. The computational complexity of the path detectors in \eqref{pathDetect} and \eqref{pathDetect_Direct} is $\mathcal{O}\left(NM^2n_p+MN^2n_p\right)$. Note that for calculating the order of computational complexity of these blocks, we use the fact that $\mathrm{trace}(\mathbf{A}\mathbf{B}\mathbf{B}^H\mathbf{A}^H)$ can be implemented with a computational complexity of order $\mathcal{O}(ABC)$, where $\mathbf{A}$ and $\mathbf{B}$ are $A\times B$ and $B\times C$ matrices. Performing the SIC in \eqref{SIC_blocked} at the $j$th iteration has a complexity of order $\mathcal{O}(j^3+j^2Mn_p)$. The number of iterations for performing JDCE is upper bounded by $T_a = K_aL_a/S$, where $L_a$ is the average number of paths between each user and RIS. Therefore, the total computational complexity of JDCE in $S$ slots is then upper bounded by 
\begin{align}
    \mathcal{O}\left(ST_a MNn_p\left(2^{B_p}+M+N\right)+S\sum_{j=1}^{T_a}(j^3+j^2Mn_p)\right).\label{compxty1}
\end{align}
\subsubsection{Data Detection}
The computational complexity for the LLR generation and the SIC steps in Sections \ref{sec_polar} and \ref{sec_polar_multi} is in the order of $\mathcal{O}((Mn_s)^3+  |\mathcal{S}_s| (Mn_s)^2)$, and the polar list decoder has a computational complexity of $\mathcal{O}( |\mathcal{S}_s| n_d\log{n_d})$. Considering $ |\mathcal{S}_s| \approx K_a/S$, the total computational complexity for the data detection part can be approximated as 
\begin{align}
   \mathcal{O}\left(S(Mn_s)^3+K_aM^2n_s^2+K_a n_d\log{n_d}\right).\label{compxty2}
\end{align}

\subsubsection{RIS Design} We now calculate the computational complexity of ASDR and AEVD. Calculating $\hat{\mathbf{R}}^{-1}$ in both algorithms has a complexity of $\mathcal{O}((Mn_s)^3+ |\mathcal{S}_s| (Mn_s)^2)$. Calculating $\mathbf{C}_i (\bar{\mathbf{w}}_c)$ and $\mathbf{C}_i^\prime (\bar{\mathbf{w}}_c)$ in \eqref{Eq_Ci} and \eqref{C_prime} have a computational complexity of $\mathcal{O}(|\mathcal{S}_s| MN^\prime n_s^2(M+N))$. Finding the $\mathbf{q}_{i,\max}$ in AEVD has a computational complexity of $\delta_{\mathrm{AEVD}}= \mathcal{O}(|\mathcal{S}_s| {N^\prime}^2)$ \cite{Lax2007Linear}. For performing SDR in ASDR algorithm, the computational complexity is $\delta_{\mathrm{ASDR}}= \mathcal{O}\left(\max(|\mathcal{S}_s|,N^\prime)^4 \sqrt{N} \log (1/\epsilon^\prime) \right)$, where $\epsilon^\prime>0$ denotes the solution accuracy \cite{Luo2010semi}. Considering $|\mathcal{S}_s|\approx K_a/S$, the total complexity of ASDR and AEVD algorithms (which is performed in $T_{\mathrm{iter}}$ iterations) in $S$ slots can be approximated as 
\begin{align}
\mathcal{O}\left(ST_{\mathrm{iter}}\left(M^3n_s^3+    |\mathcal{S}_s| M  N^\prime n_s^2(M+N) + 
 \delta_{\mathrm{alg}} \right)\right), \label{compxty3}
 \end{align}
where $\mathrm{alg} \in \{\mathrm{ASDR},\mathrm{AEVD}\}$. Looking at \eqref{compxty3}, it is clear that employing $\mathcal{C}_0$ structure significantly decreases the computational complexity of the RIS design algorithms. Moreover, the computational complexity of AEVD is lower than that of ASDR.
\section{Numerical Results}
\label{sectionSimulationResults}
In this section, we provide several numerical examples to demonstrate the performance of the proposed RIS-aided URA approach. In all the simulations, unless otherwise stated, we consider the user-BS link to be completely blocked, and we choose $\sigma_z^2=-95$dBm, $B_c=90$, $B_p=10$, $S=4$, $n_d=256$, $n_s=10$, $n_p=440$, $r=16$, $L_G=2$, $L_{R,i} =2, i = 1,...,K_a$, $M_1=M_2=N_1=N_2 = 8$, $\alpha_1=0.01$, $L_0=10^{-3}$, $\alpha_{\mathrm{PL}} = 2.3$ (for user-RIS and RIS-BS paths), and $d=\lambda/2$. The distances of each user-RIS and each user-BS paths are drawn from $d_i\in U(200m,300m)$ and $d_i\in U(250m,350m)$, respectively, and the distance is set as $d_i=100m$ for every RIS-BS path. We draw every element of $\mathbf{B}$, $\mathbf{P}$, and $\mathbf{W}_{p_s}$ from $\mathcal{CN}(0, 1)$, then each element of $\mathbf{W}_{p_s}$ is rescaled to have unit modulus, every row of $\mathbf{B}$ is normalized to have a norm of $n_s$, and rows of $\mathbf{P}$ are scaled to satisfy $\|\mathbf{W}_{p_s}\mathrm{diag}(\bar{\mathbf{p}}_i)\|^2=\|\mathbf{W}_{p_s}\mathrm{diag}(\bar{\mathbf{p}}_j)\|^2$, $\dfrac{1}{2^{B_p}}\sum_{i=1}^{2^{B_p}}\|\bar{\mathbf{p}}_i\|^2=n_p$ (for a completely blocked user-BS path), and $\|\bar{\mathbf{p}}_i\|^2=\|\bar{\mathbf{p}}_j\|^2=n_p$ (for non-blocked user-BS path) with $\bar{\mathbf{p}}_i$ being the $i$th row of $\mathbf{P}$. To construct the steering vectors of each path for the RIS in \eqref{eqs1} and \eqref{eqs3}, we randomly choose $\bar{\phi}$ and $\bar{\psi}$ from $\mathcal{T}(N_1)$ and $\mathcal{T}(N_2)$, respectively, where $\mathcal{T}(.)$ is as defined in the last paragraph of the introduction section. Similarly, the AOAs of each received path at the BS in \eqref{eqs1} and \eqref{eqs4} are randomly selected from the sets $\mathcal{T}(M_1)$ and $\mathcal{T}(M_2)$. The energy-per-bit and the PUPE of the system are defined as 
\begin{align}
    E_b/N_0 &= \dfrac{P n_T}{B}\\
    P_e &= p_{fa}+p_{md},\label{PUPE}
\end{align}
where $P$ and $n_T$ are the average per-symbol power and the total length of the transmitted signal of each user, $p_{md}$ is the probability that an active user's message is not decoded, and $p_{fa}$ is the probability that a decoded message is indeed not sent.  

\begin{figure}[t!]%[h!]
	\centering
	\includegraphics[width=1\linewidth]{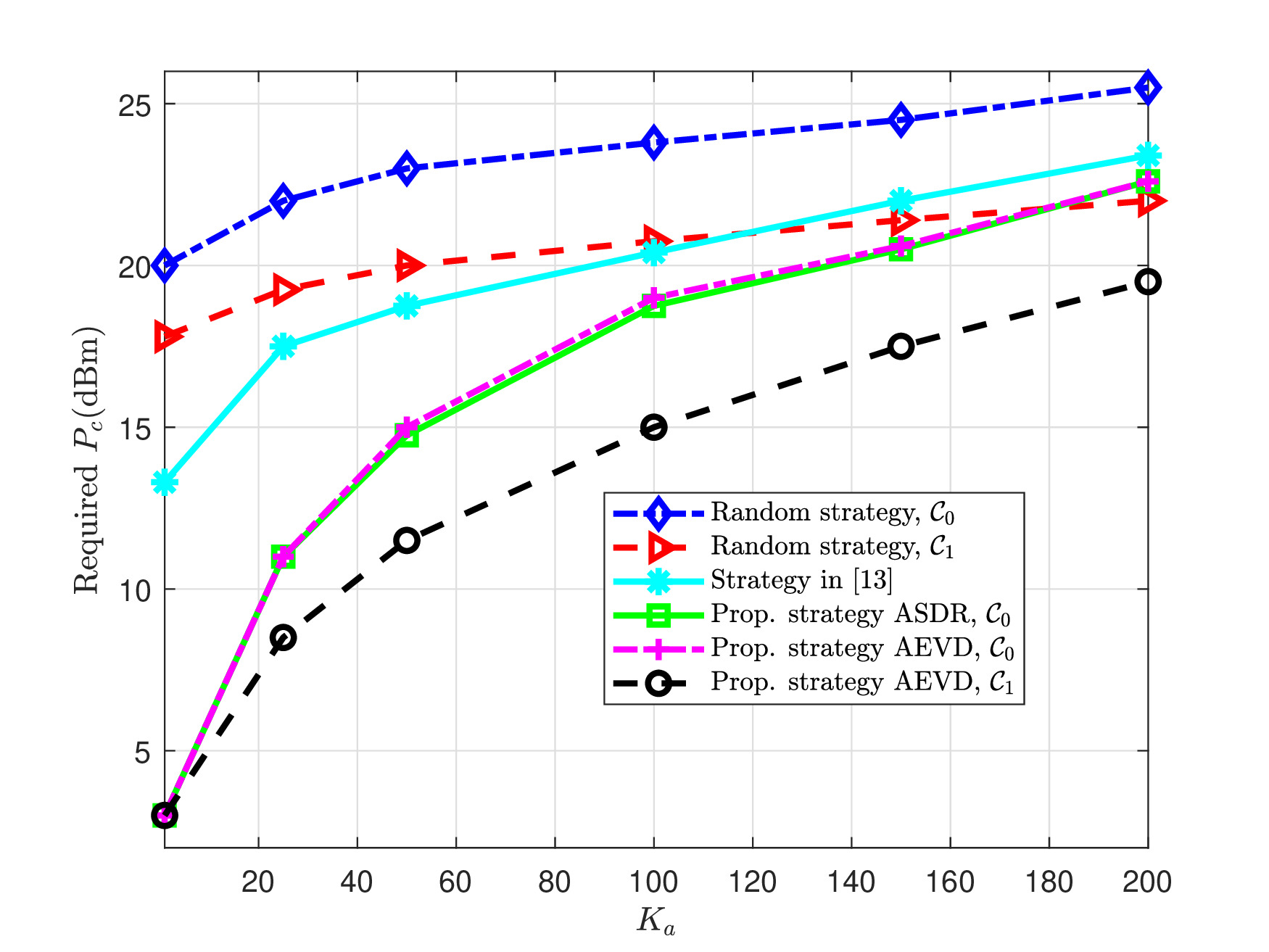}
	\caption{{\small The required $P_c$ to achieve the target PUPE of 0.1 for $n_s=10$, $n_d=256$, and different RIS phase shift strategies.}	}
	\label{Fig_beamForming}
\end{figure}

 In Fig. \ref{Fig_beamForming}, we depict the average transmit power required by the proposed scheme to achieve the target PUPE of 0.1 for different RIS design strategies and different RIS reflection algorithms defined in \eqref{W_c_structures}. Note that these results assume that the CSI is known. We can deduce from this figure that the proposed beamforming strategy performs better than the strategy employed in \cite{Shao2022reconf} and the case of randomly generated phase shifts (resulting in up to $11$dBm and $17$dBm power savings, respectively). This improvement is due to the employment of a more suitable metric for the RIS design in the proposed algorithm, where the overall phase shift matrix is obtained by minimizing the decoding error probability. This is in contrast to the RIS design algorithm in \cite{Shao2022reconf} maximizes the minimum channel gain among the active users. Besides, it is clear that the proposed AEVD performs comparably to the proposed ASVD (with $\bar{\alpha}=0.53$), while having lower computational complexity. Also, we can interpret from this figure that employing $\mathcal{C}_1$ structure (which involves varying RIS coefficient vectors across different symbol durations) provides higher accuracy, however, according to our discussion in Section. \ref{Section_COmputaional}, it suffers from a considerably higher computational complexity than the $\mathcal{C}_0$ case, where the RIS coefficient vector remains constant within a slot. Note that since we consider the direct link between each user and the BS to be completely blocked in this simulation, no communication is possible without the help of the RIS.
 
\begin{figure}[t!]%[h!]
	\centering
	\includegraphics[width=1\linewidth]{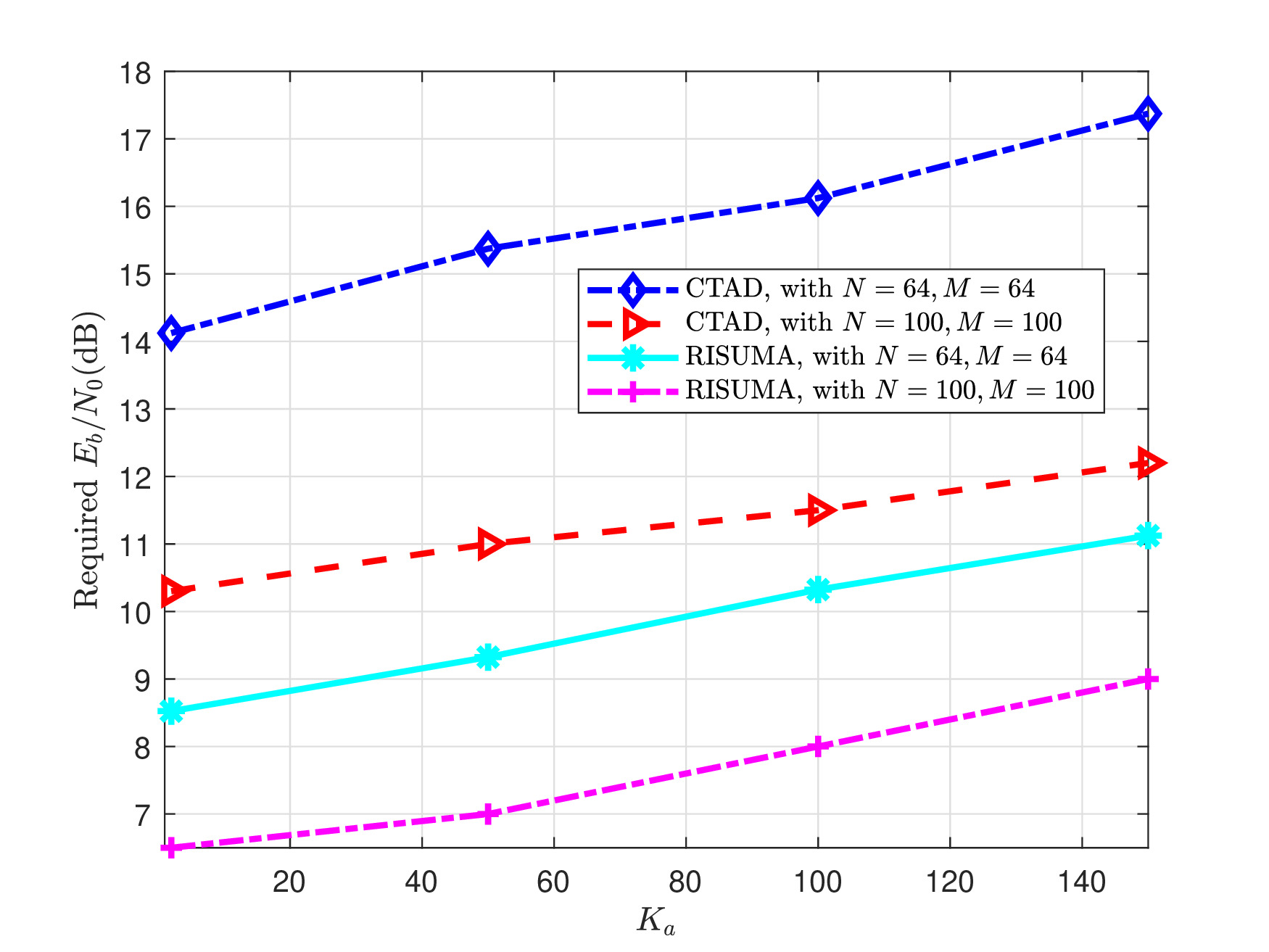}
	\caption{{\small The required $E_b/N_0$ for the proposed RISUMA and CTAD in \cite{Shao2022reconf} for achieving a target PUPE of $0.1$.}	}
	\label{FIG_RAYLEIGH}
\end{figure}

As shown in \eqref{eqs1}, RISUMA is designed based on the Saleh-Valenzuela model for the RIS-BS channel. However, to ensure a fair comparison with CTAD algorithm in \cite{Shao2022reconf}, we also consider a Rayleigh channel model, i.e., the elements of RIS-BS channel matrix are generated as $[\mathbf{G}]_{(i,j)}\sim \mathcal{CN}(0, L_0d_l^{\alpha_{\mathrm{PL}}})$, where $\mathrm{rank}(\mathbf{G})=\min(M,N)$. The performance of the RISUMA and CTAD for unknown CSI is compared in Fig. \ref{FIG_RAYLEIGH} for $n=12288$ ($S=12$, $n_d=512$, $n_s=1$, $P_c/P_p=0.5$, $n_p=512$, and AEVD for RISUMA), $B=316$, and different values of $M$ and $N$. We can see from this figure that the proposed RISUMA shows superior performance compared to CTAD. The reasons for this are: 1) RISUMA adopts a more reliable RIS design algorithm than CTAD, i.e., minimizing the decoding error of the polar code instead of maximizing the minimum channel gain of the users in CTAD, 2) RISUMA employs a powerful polar list decoder for decoding transmitted messages, while CTAD uses a dequantizer for mapping soft estimated codewords to the bit sequences, 3) RISUMA employs SIC, which is an effective block in the URA set-up \cite{ahmadi2021Unsourced,Ahmadi2023Unsourced}. We should also note that in the case of a rank-deficient $\mathbf{G}$ matrix ($\mathrm{rank}(\mathbf{G})<\min(M,N)$), there is a meaningful performance degradation in CTAD algorithm; however, the proposed RISUMA solution does not make any assumptions on the rank of this matrix, and is almost unaffected by it.

To assess the contribution of RIS to the efficiency of URA, in Fig. \ref{FIG_WithoutRIS}, we compare the performance of the proposed RISUMA approach with and without employing RIS for achieving a target PUPE of $0.1$. We assume that user-BS communication channels exist, and different path-loss exponents $\alpha_{\mathrm{PL}}=3.5, 4, 4.5, \infty$. Note that $\alpha_{\mathrm{PL}}=\infty$ corresponds to the scenario of fully obstructed user-BS channels. We set $L_{B,i}=L_{R,i} = 1, \  i=1,...,K_a$, $\bar{\alpha}=2$, $P_c/P_p=0.01$, and employ the $\mathcal{C}_0$ strategy. It is shown in Fig. \ref{FIG_WithoutRIS} that for stronger user-BS channels with path-loss exponent $\alpha_{\mathrm{PL}}=3.5$, employing RIS only slightly improves the performance of the system. However, for weaker user-BS channels (with the path-loss exponents of the user-BS channel $\alpha_{\mathrm{PL}} = 4, 4.5$ and $\infty$), employing RIS improves the performance of the URA system considerably. Hence, employing RIS is crucial for scenarios with weak user-BS channels, as expected. It is seen that the AEVD's performance is similar to that of the ASDR algorithm, while having reduced computational complexity. Also, we can interpret from this figure that, for completely blocked user-BS channels, the proposed coding schemes perform as well as the approximate achievability result given in Section \ref{SecAchResult}, particularly when $K_a$ is small.
\section{Conclusions}
\label{Conclustions}
We have proposed a RIS-aided URA scheme to enable connectivity when the direct channels of the users to the BS are blocked or significantly attenuated. The proposed scheme operates in two phases: the RIS configuration phase and the data phase. In the former, transmitted pilots are identified in the presence of heavy interference, their corresponding CSI is estimated, and RIS reflection coefficients are suitably designed by employing two different approaches. In the data phase, the receiver detects the transmitted messages using a polar list decoder, and the contributions of successfully decoded messages are removed from the received signal using SIC. We have demonstrated that the proposed scheme enhances the URA system's performance when the connections between the users and the base station are significantly degraded. Moreover, it achieves a reduction in energy consumption of up to 4 dB when compared to the state-of-the-art RIS-assisted URA algorithms. We also calculate an approximate achievability bound for the RIS-aided URA in the presence of completely blocked user-BS channels, and show that the results achieved by the practical solutions are close to the bound.


\end{document}